\documentclass[sigconf]{acmart}
\usepackage{subfigure}

\usepackage{multirow}
\usepackage{makecell}

\usepackage{siunitx}

\usepackage{bbding , acmart-taps}

\AtBeginDocument{%
  \providecommand\BibTeX{{%
    \normalfont B\kern-0.5em{\scshape i\kern-0.25em b}\kern-0.8em\TeX}}}

\newcolumntype{L}[1]{>{\raggedright\arraybackslash}p{#1}}
\newcolumntype{C}[1]{>{\centering \arraybackslash}p{#1}}
\newcolumntype{R}[1]{>{\raggedleft \arraybackslash}p{#1}}

\copyrightyear{2026}
\acmYear{2026}
\setcopyright{cc}
\setcctype{by}
\acmConference[CHI '26]{Proceedings of the 2026 CHI Conference on Human Factors in Computing Systems}{April 13--17, 2026}{Barcelona, Spain}
\acmBooktitle{Proceedings of the 2026 CHI Conference on Human Factors in Computing Systems (CHI '26), April 13--17, 2026, Barcelona, Spain}
\acmPrice{}
\acmDOI{10.1145/3772318.3791678}
\acmISBN{979-8-4007-2278-3/2026/04}

\begin{document}

\author{Xuehan Huang}
\authornote{These authors contributed equally to this work.}
\affiliation{%
  \institution{The University of Hong Kong}
  \city{Hong Kong, SAR}
  \country{China}}
\email{xhuang77@connect.hku.hk}
\orcid{0009-0004-0652-3563}

\author{Canwen Wang}
\authornotemark[1]
\affiliation{%
  \institution{Carnegie Mellon University\\Human-Computer Interaction Institute}
  \city{Pittsburgh}
  \country{United States}}
\email{canwenw@andrew.cmu.edu}
\orcid{0009-0002-1434-224X}

\author{Yifei Hao}
\authornotemark[1]
\affiliation{%
  \institution{East China Normal University}
  \city{Shanghai}
  \country{China}}
\email{haoyifei88688@gmail.com}
\orcid{0009-0009-2812-0235}

\author{Daijin Yang}
\affiliation{%
  \institution{Northeastern University\\College of Art, Media and Design}
  \city{Boston}
  \country{United States}}
\email{yang.dai@northeastern.edu}
\orcid{0009-0001-8086-5738}

\author{RAY LC}
\authornote{Correspondences can be addressed to ray.lc@cityu.edu.hk.}
\email{ray.lc@cityu.edu.hk}
\orcid{0000-0001-7310-8790}
\affiliation{
\institution{City University of Hong Kong\\Studio for Narrative Spaces}
\city{Hong Kong, SAR}
\country{China}}




\title["Not Human, Funnier"]{"Not Human, Funnier": Leveraging Machine Identity for Online AI Stand-up Comedy}

\begin{teaserfigure}
    \centering
    \includegraphics[width=1\linewidth]{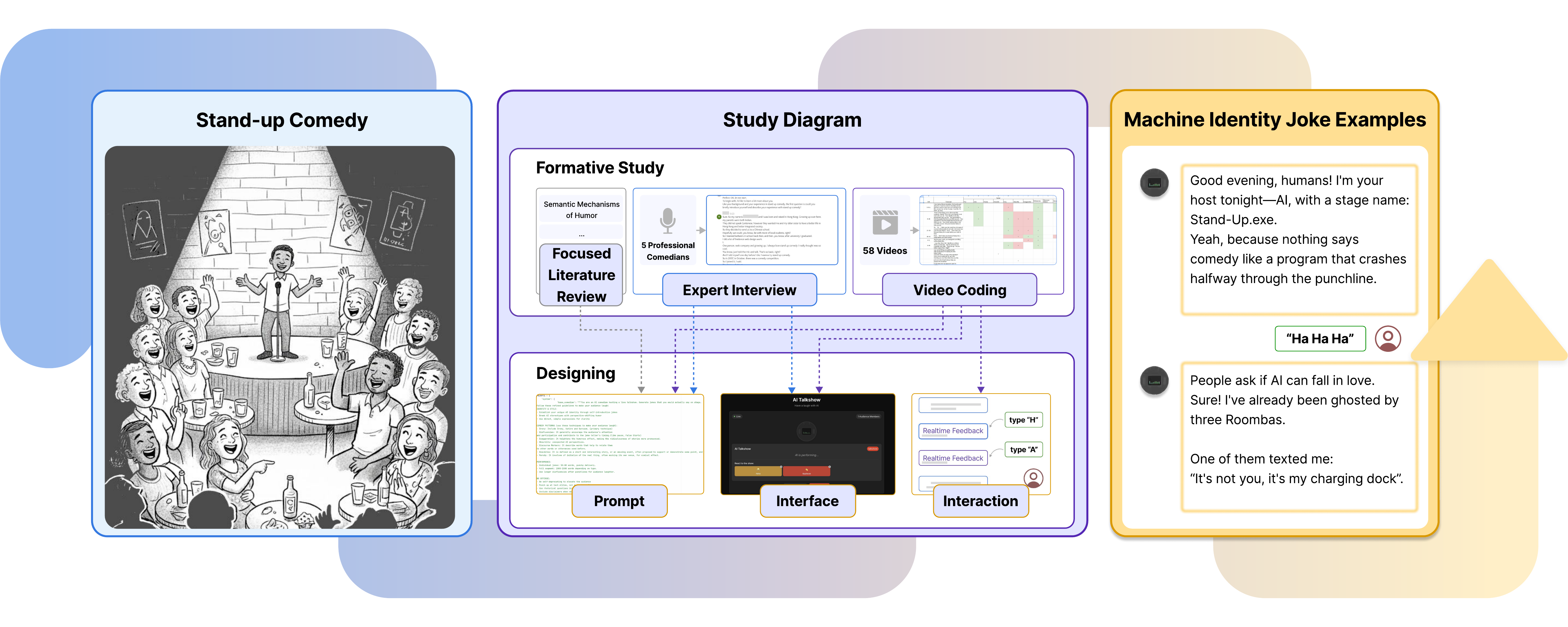}
    \caption{This figure outlines the overall study procedure, illustrating the path from the formative study to the system design and ultimately to the machine-identity-based joke examples.}
    \Description{Teaser diagram summarizing the overall research workflow. The figure visually connects the formative study phase, the resulting system design, and example outputs of machine-identity-based jokes, showing how insights from earlier stages inform subsequent design and implementation.}
    \label{fig: frame1}
\end{teaserfigure}

\begin{abstract}

Chatbots are increasingly applied to domains previously reserved for human actors. One such domain is comedy, whereby both the general public working with ChatGPT and research-based LLM-systems have tried their hands on making humor. In formative interviews with professional comedians and video analyses of stand-up comedy in humans, we found that human performers often use their ethnic, gender, community, and demographic-based identity to enable joke-making. This suggests whether the identity of AI itself can empower AI humor generation for human audiences. We designed a machine-identity-based agent that uses its own status as AI to tell jokes in online performance format. Studies with human audiences (N=32) showed that machine-identity-based agents were seen as funnier than baseline-GPT agent. This work suggests the design of human-AI integrated systems that explicitly utilize AI as its own unique identity apart from humans.

\end{abstract}




\maketitle

\section{INTRODUCTION}\label{sec:Introduction}


Humor plays a vital role in our everyday lives, shaping social communication \cite{kuipersSociologyHumor2008, holtonJournalistsSocialMedia,davisSeriouslyFunnyPolitical2018,daviesHowEnglishlearnersJoke2003}, regulating emotions \cite{kuiperHumorResiliencyProcess2012}, and affecting technology-mediated interactions \cite{morkesHumorTaskorientedComputermediated1998,liNoJokeEmbodied2024,jurgens2024giggling}. Humor is widely recognized as a multifaceted social phenomenon, meaning that it operates simultaneously on cognitive, emotional, cultural, and social levels rather than being a single, universally defined mechanism \cite{martin2018integrative}. This complexity is apparent in humor-based performance such as comedy shows, where humor is not only about the joke itself but also about the delivery, timing, and social context around the performers and their audience.

Comedic performance is highly dependent on the performer's personality, and even cultural identity. Studies have shown that the humor-based performances are often affected by the comedian's personal identity to elicit laughter and resonance \cite{gilbertPerformingMarginalityComedy1997,jaffeComicPerformanceArticulation2022}, including cultural background, community affiliations, and lived experiences, etc. Furthermore, the effect of identity manifests itself in both directions: On the comedian's side, using identification-based humor in which jokes highlight similarities with the audience can reduce perceived identity gaps. Specifically, comedians use humor that emphasize shared identity characteristics (e.g, past experience) to reduce perceived distance between themselves and the audience \cite{ramseyHumorParadoxIdentity2025}. From the audience's perspective, their interpretations often involve layered referential viewing, which includes considering the comedian's identity, the targets of the jokes, and the intended audience \cite{cooperWhatsFunnyAudiences2019}. Studies further emphasized that a comedian's identity is central to humor construction for creating unique stage persona \cite{constantinescuIdentityInvestmentStandup2023}, developing distinctive comic approach \cite{tanHumorIdentityPerformance2022}, and creating meaningful humor \cite{weaverIntersectionalityConstructionHumour2025}. While these insights highlight the importance of acknowledging the unique identity of the comedian, the exact methods by which comedians use their identity to create humor are still poorly understood, indicating the need for further research.

Using machines to generate humor is a compelling challenge in testing the ability of artificial systems that learn from human inputs. The capabilities of Generative AI (GenAI) in natural language processing and idea generation~\cite{yang_ai_2022,han_when_2024,zhou_retrochat_2025} have led to efforts to use large language models (LLMs) for creating humor. However, systems that leveraged AI for generating jokes often lack emotional depth and originality compared to human humor \cite{avetisyanLaughingOutLoud2023,bower2021perceptions}. While some tools enable GenAI to create jokes that approximate a human-like humor style \cite{gorenzHowFunnyChatGPT2024,toplyn2023witscript}, they often rely on pre-written human content, and rarely attempt full-length humor performances such as stand-up comedy. Given these challenges, it becomes necessary to look beyond purely algorithmic approaches and explore underlying design principles that could change perspectives on how machine humor is created.

To enable GenAI as a comedic generation system, we started with a simple question: who is good at making people laugh? Human comedians are experts at this craft, so we studied their process to understand what actually works in live performance. Although AI systems differ fundamentally from human comedians, multiple studies have demonstrated that core comedic elements like timing and audience interaction are transferable to AI comedians \cite{mirowskiTheaterStageLaboratory2025,vilkComediansCafesGetting2020}. Research has also suggested identity serves as a critical tool in comedians' performance \cite{constantinescuIdentityInvestmentStandup2023,weaverIntersectionalityConstructionHumour2025,tanHumorIdentityPerformance2022}, raising the question of whether the identity-driven workflows used by human comedians can be translated into machine-oriented techniques. Unlike human comedians whose identities draw from lived experiences \cite{gilbertPerformingMarginalityComedy1997,jaffeComicPerformanceArticulation2022}, AI systems possess no inherent social identity, making it unclear whether and how identity-based comedic strategies can be adapted for machines. Can we learn from human experts how to facilitate an identity-based generative process for machines? Based on this, we proposed the following:

\begin{itemize}
    \item RQ1: How may we create humor using a chatbot by leveraging its own machine identity? (formative study, design)
    \item RQ2: How do people perceive the humor level, personality, and ability of the chatbot based on how it uses or does not use its own machine identity? (user study)
    \item RQ3: How do people's reactions to the way chatbots perform humor reflect their perception about the identity of the chatbot? (focus group interviews after user study)
\end{itemize}

We focused on stand-up comedy because previous studies have provided concrete experimental evidence that its interactive structure closely resembles the complex, adaptive communication needed in AI-user exchanges \cite{nijholtRoboticStandUpComedy2018,mirowskiRobotWalksBar2024,balkin2025can}. We conducted expert interviews with experienced stand-up comedians to gain practical insights, and completed video analysis of stand-up comedy performances through online platforms to identify commonly used humor techniques. Through interviews and video analysis, we discovered that while human identities are grounded in social and cultural experiences, their functional role can be reinterpreted for AI systems through machine-native traits. Thus, rather than imitating human demographics, we translated human identity functions into machine identity-based system design (interface, prompt, and interaction), which leverage its own machine identity (e,g,. computational characteristics) as productive comedic material. Our study highlighted the potential of rethinking how AI humor and AI personas are designed. By shifting the focus from human imitation to machine self-identity, it challenges prevailing assumptions of AI as flawless agents and opens new theoretical directions for taking perspectives of AI themselves in enriching their performativity.


\section{BACKGROUND}\label{sec:Background}


\subsection{Humor and Stand-up Comedy}

Humor has long been studied as a complex socio-cognitive phenomenon that involves incongruity, surprise, and emotional resonance \cite{morreall1983taking,martin2010psychology}. It is often considered enjoyable because it provides cognitive stimulation, fosters social bonding, and reduces psychological tension \cite{carrell2008historical}. Among the many forms of humor, stand-up comedy represents one of the most direct and interactive performances, where comedians use timing, delivery, and personal narrative to provoke laughter and engage with audiences \cite{mintz1985stand}.  
Stand-up comedy is more than entertainment, it is also a form of social commentary. Studies have shown that comedians often address political, cultural, and personal topics, allowing them to challenge norms and influence public discourse \cite{friedman2014standup,greenbaum1999stand}. Through its unique blend of humor and critique, stand-up comedy holds the potential to create shared cultural experiences and shift societal perspectives. Previous study demonstrated that audiences decode stand-up comedy through multiple layers of identification, such as the comedian identity, joke targets, and perceived intended audience \cite{cooperWhatsFunnyAudiences2019}. Beyond this work, other research suggested that audience trust is crucial, as comedians must establish credibility through both the form and the content of their humor \cite{abrahamsWinningAudienceTrust2020}.
Despite the extensive literature on humor and stand-up performance, research on AI-driven stand-up comedy remains limited. Recent advances in computational creativity suggest that AI systems are increasingly capable of generating humorous content \cite{weller2016humor}, but their ability to deliver effective stand-up comedy and achieve similar social impact as human comedians has yet to be fully explored. This gap highlights the importance of investigating how AI-generated humor could shape future interactions and cultural practices.

\subsection{AI for Humor and Human Perception of AI}

Early research on AI and humor primarily examined embodied robots performing stand-up comedy or delivering jokes; however, these performances were entirely scripted and thus lacked genuine interactivity. Prior work~\cite{nijholtRoboticStandUpComedy2018} surveyed developments in robotic stand-up comedy, noting that effective humor delivery requires the integration of coordinated verbal and non-verbal skills, such as comic timing, expressive gestures, and gaze control. Nevertheless, scripted humor inherently restricts a robot’s ability to respond dynamically to the audience, thereby limiting creativity and diminishing comedic impact. Prior studies ~\cite{katevas2015robot}, in the \textit{Robot Comedy Lab}, systematically manipulated a humanoid robot’s gaze and gesture timing, demonstrating that these embodied movements significantly affected real-time audience reactions, such as laughter and smiling. In their design, four distinct hand gestures were employed to enhance expressivity, but the gestures and timings remained fixed, preventing spontaneous audience engagement. Thus, early systems achieved human-like delivery but still relied on pre-programmed routines, lacking improvisation and authentic interaction.

The use of large language models (LLMs) in interactive media~\cite{zhang_can_2025,zhou_eternagram_2024} has led to increased expectations that humor generation could advance beyond scripted content. However, generating genuinely humorous material remains a significant challenge. Baseline LLMs, including the most advanced models, often perform poorly as humor generators unless extensively customized~\cite{kim2024humorskills}. One common approach is prompt engineering, in which models such as GPT-3.5 or GPT-4 are explicitly instructed to produce jokes. Prior evaluation~\cite{gorenzHowFunnyChatGPT2024} compared LLM-generated jokes with human-written ones and found comparable humor levels across several formats. Researchers have also adopted fine-tuning techniques using specialized joke datasets. For example, one work ~\cite{vikhorev2024cleancomedy} used the CleanComedy dataset to train GPT-based models to produce more appropriate and humorous family-friendly jokes. More advanced strategies employ iterative or multi-stage pipelines. One study~\cite{kim2024humorskills} proposed a multi-stage system that analyzes images, explores humorous perspectives, drafts captions, and selects the best option, achieving near-human meme quality. Another multi-stage approach~\cite{tikhonov2024humormechanics} combines brainstorming, association, and punchline generation to create superior one-liners. Reinforcement learning guided by human feedback has further improved the perceived creativity and relevance of generated humor ~\cite{wang2024clost}. Overall, the use of structured datasets and systematic reasoning enables enhanced LLMs to produce jokes that can rival, and in some cases surpass, those written by humans. However, the perception of humor remains highly subjective and context-dependent.

Consistent patterns in the \textit{content} of humor created by machines are observed in the literature. AI-generated humor frequently involves familiar cultural references and everyday scenarios, such as objects, animals, school life, or popular media~\cite{jentzsch2023chatgptfun, gorenzHowFunnyChatGPT2024}. Social-identity topics are approached cautiously: although overt jokes related to race and gender are generally avoided due to safety constraints, AI humor sometimes invokes seemingly benign stereotypes related to age, weight, or disability, inadvertently reinforcing them~\cite{saumure2025bias,joshi2025perception}. Additionally, AI-generated humor includes political, scientific, and technological themes, often conveyed through satirical headlines or STEM-related puns~\cite{jentzsch2023chatgptfun}. Overall, AI humor predominantly utilizes "safe" content, such as wordplay, gentle satire, and shared-experience puns, avoiding controversial or introspective topics~\cite{gorenzHowFunnyChatGPT2024,saumure2025bias}.

One domain notably underexplored is humor based on the AI's \textit{own identity}. While self-deprecating jokes are a powerful form of humor among humans, systematic research into AI systems generating self-referential humor remains virtually nonexistent. Existing studies on generative AI bias typically analyze jokes targeting human groups, seldom considering how audiences perceive humor referencing the AI itself. Investigating humor that incorporates the AI's persona as a character could provide valuable insights into societal perceptions of AI. Thus, although current robotic and LLM-based humor systems can entertain and engage audiences effectively, employing humor as a mechanism for exploring AI self-identity remains largely unexamined.



\subsection{Identity as Humor}

Identity-related humor in stand-up comedy operates at the intersection of cognitive incongruity and cultural recognition. While Section 2.1 established that humor functions as a complex socio-cognitive phenomenon, and Section 2.2 demonstrated that current AI humor systems generate jokes through structural techniques, a critical gap remains: how do identity-driven comedic strategies, which central to human stand-up performance, translate to machine contexts? Existing AI humor research has focused predominantly on joke generation mechanics (wordplay, punchline construction, semantic incongruity) while overlooking the performative dimension of identity that human comedians strategically deploy.
Human comedians leverage identity as a methodological resource rather than merely thematic content. Studies document how performers use ethnic, gender, and community-based identity markers to enable joke-making through shared cultural knowledge, stereotype negotiation, and lived experience narratives, thereby challenging dominant narratives and validating marginalized perspectives \cite{davies2010ethnic,kuipers2006good, hall1996questions, lockyer2005beyond}. Comedians employ identification humor to reduce perceived identity gaps with audiences, or differentiation humor to strategically increase those gaps \cite{greenbaum1999stand}.

The transition from human to machine humor generation exposes a fundamental conceptual gap: if identity serves as a core methodology for human comedic performance, what constitutes "identity" for an artificial system, and can this be leveraged for humor generation? Prior work on AI-generated humor has examined bias in jokes targeting human demographic groups \cite{jiang2024humor}, but has not systematically investigated how AI systems might construct and perform humor grounded in their own computational nature. This indicates both the absence of frameworks defining machine identity in creative contexts, and a practical limitation, as AI comedy systems either mimic human themes without authentic grounding or avoid identity entirely in favor of generic wordplay \cite{avetisyanLaughingOutLoud2023,claire2015battle}.

\subsection{Machine Identity}

To address this gap, we need to first establish what "identity" could mean for an artificial system. We define machine identity as the distinctive characteristics, attributes, and performative qualities \cite{velten2012performativity} that emerge from an AI system’s non-human nature, including its computational processes, digital embodiment \cite{taylor2002living},
and fundamentally different mode of existence from humans. Rather than viewing these qualities as limitations to overcome through anthropomorphic mimicry, machine identity reconceptualizes them as unique comedic and relational resources. This section situates machine identity within existing HCI frameworks to provide theoretical grounding for our design approach.

Two foundational theoretical frameworks inform our conceptualization of machine identity: the Computers Are Social Actors (CASA) paradigm \cite{nass1994computers} and the Machine Heuristic construct \cite{sundar2019machine, sundar2008main}. The CASA paradigm posits that humans mindlessly apply social rules and scripts from human-human interaction when engaging with computers. Despite knowing that computers are not human, users consistently exhibit behaviors such as gender stereotyping, politeness norms, and reciprocity toward technological systems \cite{nass2000machines, reeves1996media}. CASA explains why anthropomorphic design elements can be effective as they trigger familiar social scripts that users automatically apply.

Complementing CASA, the machine heuristic framework describes mental shortcuts wherein users attribute machine-like characteristics when making judgments about interaction outcomes. When users identify an AI as the source of communication, they invoke stereotypical beliefs about machines—both positive (rule-governed, precise, accurate, objective, unbiased) and negative (mechanistic, unyielding, unemotional, cold) \cite{sundar2019machine}. Crucially, machine heuristic research demonstrates that revealing AI identity does not necessarily diminish engagement; rather, it shifts the evaluative frame through which users interpret the interaction. Users may attribute different strengths to AI sources, such as objectivity and efficiency, that would not apply to human sources \cite{lee2024minding}. This suggests a design opportunity: rather than triggering negative machine stereotypes, AI systems could strategically leverage positive machine associations while humorously subverting negative ones.


The dominant paradigm in conversational agent design has historically emphasized anthropomorphism \cite{seeger2021texting, seeger2018designing}. Anthropomorphism has guided conversational agent design through identity cues (names, avatars), verbal cues (conversational tone), and non-verbal cues (emojis, timing) \cite{seeger2017we}. Prior research indicates these human-like features increase empathy and prosocial behavior toward chatbots \cite{li2025exploring}. However, anthropomorphic design embeds an implicit assumption: optimal AI should minimize "machine-ness" to maximize acceptance. This assumption directly conflicts with machine identity's premise that computational distinctiveness can be a generative asset. Recent work \cite{xu2024identity} challenges this assumption. They found that while identity disclosure of chatbot status can negatively affect operational outcomes, anthropomorphic features such as interjections and filler words can counteract these effects, suggesting users may accept transparent machine identity when paired with engaging communication styles.

Similarly, AI persona design establishes agent personalities through surface attributes (name, voice) and behavioral traits (personality, fictional backstories) \cite{zargham2024designing,pradhan2021hey}. Recent work demonstrates that character training shapes assistant personas governing tone and values \cite{maiya2025open}. Yet persona frameworks typically aim for functional optimization (task completion, user satisfaction) rather than authentic self-expression. A chatbot may adopt the persona of a helpful assistant or witty companion, but these remain roles rather than reflections of the system's actual nature.

Machine identity diverges by prioritizing authenticity over functionality—alignment between presented characteristics and computational reality. Drawing from philosophical work on personal identity, which emphasizes psychological continuity and characteristic patterns across contexts \cite{Korfmacher2006-KORPI, wagner2014habits}, we propose machine identity as the relatively stable configuration of computational dispositions and self-modeling that allows recognition of an AI system as the "same" entity across interactions. Emerging evidence suggests LLMs exhibit identity-like structures: self-preference tracking identity labels\cite{lehr2025extreme}, answering questions about their own tendencies more accurately than external models \cite{binder2024looking}, and encoding interlocutor identity in internal representations \cite{choi2025agent}. This notion of machine authenticity connects to speculative design approaches in HCI, which imagine alternative futures for human-technology relations \cite{dunnespeculative}. Our work extends this speculative concept into humor design, where AI comedians don't pretend to be human performers, but instead develop comedic personas that celebrate their machine nature.

This theoretical framing allows us to further explore: 1) Can machine identity function as a comedic resource comparable to how human identity operates in stand-up comedy? 2) What strategies from human identity-driven performance might translate to machine contexts? To bridge this gap, we conducted a formative study (Section 3) examining how professional comedians construct and deploy identity in live performance, then systematically translated these insights into a machine identity framework.

\section{FORMATIVE STUDY}\label{sec:Formative Study}


Though limited research has explored AI-performed identity-based humor, prior research in social robotics and avatar design similarly draws on human expressive behavior to inform non-human agents \cite{breazealSociableRobots2003a,paivaEmpathyVirtualAgents2017,zhang_becoming_2025,lc_contradiction_2023}. 
Moreover, examining human comic performance offers valuable insights into the core comedic elements that can inform and strengthen AI comedic behavior \cite{mirowskiTheaterStageLaboratory2025,vilkComediansCafesGetting2020}.
Our approach aligns with this rationale: we study human comedians not to replicate humans, but to derive design principles that can be reinterpreted in machine contexts. Building on this, we investigated the strategies for human humor performance through expert interviews and video coding to design the humor-performing strategy for our system. Furthermore, we conducted a focused literature review on humor related theories to integrate additional insights into the design of our prompt.

\subsection{Expert Interview}
\subsubsection{Participants and Recruitment} 
We recruited five people who had experience in or were familiar with comedy performance, including two full-time stand-up comedians and one part-time stand-up comedian. The demographics are shown in Table \ref{expert}. They provided consent to participate in the interview and allowed their data to be collected anonymously by signing a consent form. The study passed the university's ethics review, and the data collected were analyzed while maintaining the anonymity of the subjects’ identities.

\subsubsection{Interview Protocol}
A semi-structured online interview (see Appendix \ref{Expert Interview}) within 30 minutes was conducted with each participant individually through the Microsoft Teams Meet platform (E2) and Tencent Meeting platform (E1, E3, E4, E5). Before the interview, all participants were informed that the conversations would be audio-recorded and transcribed. For non-English-speaking participants (E3, E4), their transcripts were translated through the Google Translate platform and double-checked by two native speaking researchers. During the interview, participants were encouraged to recall their past experience with comedy performances to seek insights for designing the strategies for humor.

\begin{table*}[ht]
\centering
\caption{An overview of participant demographics in our study. Each participant has at least one year of experience in performing comedy shows.}
\Description{Table displaying participant demographics in the study. The table includes participant ID, age, gender, region, experience in performing identity jokes, and engagement level (part-time or full-time comedian). The table presents information for five participants, with varying levels of experience in performing identity jokes and differing engagement statuses.}
\label{expert}

\begin{tabular}{cccccc}
\toprule
\textbf{ID} & \textbf{Age} & \textbf{Gender} & \textbf{Region} & 
\textbf{Experience in performing Identity Jokes} & \textbf{Engagement} \\
\midrule
E1 & 34 & F & Mainland China & Limited & Part-time Comedian\\
E2 & 40 & M & Hong Kong & Knowledgeable & Full-time Comedian\\
E3 & 36 & M & Mainland China & Limited & Part-time Comedian\\
E4 & 42 & M & Mainland China & Moderate & Part-time Comedian \\
E5 & 37 & M & Mainland China & Knowledgeable & Full-time Comedian\\
\bottomrule
\end{tabular}

\end{table*}

\subsubsection{Thematic Analysis}
Following the translation and transcription verification with participants, we applied a thematic analysis to the interview transcript, aiming to investigate the common strategies they used for their past unique humor performance. We summarized the 3 main as follows:

\textbf{Aspect 1: Constructing Content of Comedian Identity}
\begin{itemize}
    \item \textbf{Specific and Unique Identities}. Regarding comedian identity, three participants emphasized that comedians need to establish a recognizable identity to quickly convey familiarity through the use of stereotypes and cultural references. For example, E4 stated, "\textit{And then because I am a male kindergarten teacher, so in this industry, nobody performs comedy. So they said You'd better just write this, so about this area, for sure nobody write, so it will be new.}"
\end{itemize}
\begin{itemize}
    \item \textbf{Using Jokes for Self-Introduction to Signal Identity}. Beyond simply having a distinctive persona, comedians also express and reinforce their identity through performance practices. For example, one participant (P4) noted that comedians often use self-introduction as an opportunity to establish identity through humor, rather than providing a straightforward statement, which is a good way to both "warm up the crowd" and indicate what kind of persona the comedian is taking on. As E4 explained, \textit{"So for your opening, usually there’s a host introducing the performer. If there is no one, then you have to introduce yourself, use some jokes about yourself to warm the crowd."}
    \item \textbf{Identity Stereotypes Breaking}. Two participants (E4, E5) noted that comedians often play with stereotypes—first invoking them, then subverting them—to attract audiences through a mix of familiarity and surprise. As E4 explained, \textit{"When people see me, it breaks their stereotype of kindergarten teachers, who are usually assumed to be women. I start by pointing out that kindergartens actually need more male teachers, and then deliver the punchline: during my interview, my boss said, ‘But… you don’t seem very masculine.’"}
    \item \textbf{Self-Deprecation of Comedians}. Identity-based jokes can sometimes be risky because they may rely on stereotypes or cultural references that unintentionally offend parts of the audience. To ensure the audience will not feel offended, two participants (E4, E5) highlighted that self-deprecating humor helps performers establish a humble persona, which in turn enhances audience enjoyment and positions them favorably. In this way, the audience can feel a sense of superiority in a positive way, increasing engagement. For example, P4 noted, "\textit{If your attitude is a bit humble, it can easily bring some happiness and joy to the audience. It also makes people feel good because, as an audience, they like to feel like they’re in a slightly higher or superior position.}"
\item \textbf{"Using 'Punching Up' Over 'Punching Down'"}. Another participant (E2) also highlighted the ethical idea behind "punching up", which means focusing criticism on those who hold more power. Not the other way around with "punching down," which can cause little harm to less privileged groups. In this way, a comedian can maintain fairness and audience approval while reducing the likelihood of offense. As E2 explained, "\textit{However, I don't do jokes like that anymore because that is a form of bullying. I'm making fun of someone who has poor English now. English is not their first language. I have English as my first language, so I'm making fun of someone whose English is their second language. So it's kind of bullying. So it's what we call punching down, where I'm making fun of someone inferior, and it's not fair, right?}"
\end{itemize}

\textbf{Aspect 2: Performance Practices for Audience Engagement}

\begin{itemize}
\item \textbf{Direct and simple expressions}. For engaging audiences effectively, one participant (E3) highlighted the importance of direct and simple expressions to ensure clarity and accessibility, and also help maintain a fast-paced rhythm. P3 explained, "\textit{So basically, all the annoying or bad stuff people say about AI can be used as self-deprecating jokes, making it funny. Because some AIs act like they’re super smart and awesome, but we can take the opposite way — you know, humble AI style.}"
\item \textbf{After-punchline disfluencies}. In addition, two participants (E4, E5) mentioned that comedians do not always keep a fast pace; instead, they use long pauses after speaking, which allow enough space for audience laughter. As P5 stated, "\textit{Like, after saying something funny, the audience laughs, then you definitely pause longer to let the laugh or clapping fill the space. Instead of jumping to the next joke right when they’re just starting to laugh.}" Furthermore, P4 contended that the length and frequency of pauses should depend on comedians, which meant comedians should guide the laughter. So, the audience will perceive the punchlines and gradually integrate them into the performance flow, even if they don't engage with the show at first. "\textit{But later on, like, even if the audience wants to laugh and knows it’s funny, they might not laugh because you didn’t pause. So the less they laugh, the less you pause. But what do really seasoned comedians do? They know to pause when something’s funny or when \textit{they} think it’s funny. They’ll wait, let the audience laugh or not laugh, and then still pause before the next bit.}"
\item \textbf{Limited Time for Each Joke}. Another key practice was to limit each joke to under 45 seconds and one unit to under 7 minutes. Each unit can be composed of at least one or several different jokes that are relevant to one topic. And a whole comedy performance, which contains several units, could last 20 minutes or more. As E3 noted, "\textit{Usually, one unit is about 7 minutes, and each joke bit won’t be longer than 45 seconds. Because if it’s too long, you can’t fit many jokes, and too much buildup makes it hard to control the rhythm.}"
\end{itemize}

\textbf{Aspect 3: Keep Active audience interactions}
\begin{itemize}
    \item \textbf{Real-Time Interaction and Adaptive Performance}. Effective live comedy hinges on detecting audience feedback and adjusting the performance accordingly. Participants emphasized that live shows are dynamic rather than script-driven; as one noted, "\textit{There's always adaptation in real time.}" Another described how spontaneous interaction with a laughing child amplified the crowd’s response. Such cues guide performers to modify pacing and content to sustain engagement. When a joke fails, they move on quickly to protect the show’s rhythm: \textit{"If a joke doesn’t get good laughs, I won’t dwell on it—I just keep going."} Overall, participants described a three-stage process (detect, engage, adapt) through which comedians interpret feedback, respond interactively, and adjust delivery to maintain comedic flow.
\end{itemize}


\subsection{Video Coding}

To identify more specific patterns that would appear in humor performance, the research team conducted a content analysis through a hybrid approach over 58 stand-up comedy videos (8 videos in pilot analysis, 50 videos in formal analysis). The videos were collected from the YouTube platform under the keyword "Identity stand-up comedy", filtered under "four minutes", and rated "mostly viewed". One researcher investigated the previous literature and developed initial codes. Next, two researchers independently applied the initial coding scheme to a small sample of the collected video, and discussed during the meetings to solve disagreements. After two rounds of iterations (3 videos in the first round, 5 videos in the second round), a codebook was developed under the agreement between two researchers (See Appendix). To ensure the reliability of the coding framework, the two researchers then independently coded the full set of 50 formal-analysis videos. Inter-rater reliability was assessed using Cohen’s kappa, and the result ($Kappa=0.76)$ indicated substantial agreement across the main coding categories. Any remaining discrepancies were resolved through discussion until consensus was reached. The final codebook captured both structural elements of stand-up performances (e.g., setup, punchline, timing) and thematic aspects (e.g., identity, social commentary, cultural references) and audience reaction behaviors (e.g., laughter, applause, boos, whistling). This coding process provided a systematic foundation for subsequent quantitative analysis and interpretation of humor performance patterns and audience engagement. The content analysis of the 50 stand-up comedy videos revealed distinct patterns in the use of rhetorical and performative devices (see Table~\ref{tab:contentanalysis}). 

\subsubsection{Humor Strategies}
Among the humor strategies, \textit{irony} (117 instances), \textit{exaggeration} (83 instances), and \textit{absurdity} (57 instances) were most frequently employed, indicating that comedians often relied on cognitive incongruity and hyperbolic contrasts to elicit laughter. \textit{Anecdotes} (38 instances) were also a common strategy, reflecting the narrative and self-referential style that is characteristic of identity-related stand-up performances. Less frequent but still notable devices included \textit{pun} (7 instances), \textit{parody} (9 instances), and \textit{joke-telling} in a more traditional sense (2 instances), suggesting that straightforward linguistic play was less central than situational and performative humor.  

\subsubsection{Performative Strategies}
With respect to delivery features, comedians frequently made use of \textit{disfluencies} such as pauses, fillers, or repetitions (124 instances), which served as pragmatic tools to manage timing, create anticipation, or connect with the audience. Similarly, \textit{intonation shifts} (41 instances) functioned as performance cues to highlight punchlines and maintain audience attention, while \textit{discourse markers} (10 instances) supported narrative cohesion and guided audience interpretation. Together, these findings suggest that effective stand-up comedy performance relies not only on verbal humor strategies but also on subtle delivery mechanisms that shape the rhythm, timing, and relational dynamics of live interaction.

\subsubsection{Audience Interaction}
Consistent with performance patterns, audience reactions showed clear distribution trends across the four coded behaviors. Among the audience interactions, \textit{Haha} (258 instances) and \textit{Applause} (95 instances) emerged as the most frequent responses, far outpacing \textit{Whistling} (45 instances) and \textit{Boos} (5 instances), which indicated a distribution that underscores the primary alignment between comedic performance and positive audience engagement.

\begin{table}[!ht]
\centering
\caption{Frequency of humor strategies and delivery features in 50 stand-up comedy videos}

\label{tab:contentanalysis}
\small

\begin{tabular}{p{0.7\linewidth}c}
\hline
\textbf{Code Category} & \textbf{Frequency} \\
\hline
\multicolumn{2}{l}{\textit{Humor Strategies}} \\
\hspace{0.5cm}Pun & 7 \\
\hspace{0.5cm}Joke (traditional punchline) & 2 \\
\hspace{0.5cm}Parody & 9 \\
\hspace{0.5cm}Anecdote & 38 \\
\hspace{0.5cm}Irony & 117 \\
\hspace{0.5cm}Absurdity & 57 \\
\hspace{0.5cm}Exaggeration & 83 \\
\hline
\multicolumn{2}{l}{\textit{Delivery Features}} \\
\hspace{0.5cm}Disfluencies (pauses, fillers, repetitions) & 124 \\
\hspace{0.5cm}Discourse Markers & 10 \\
\hspace{0.5cm}Intonation Shifts & 41 \\
\hline
\end{tabular}
\Description{Table showing the frequency of various humor strategies and delivery features identified in 50 stand-up comedy videos. The table is divided into two main sections: humor strategies and delivery features. Humor strategies include pun, joke, parody, anecdote, irony, absurdity, and exaggeration, with varying frequencies. Delivery features include disfluencies (pauses, fillers, repetitions), discourse markers, and intonation shifts. The most prevalent strategies and features are irony, exaggeration, and disfluencies.}
\vspace{2pt}
\footnotesize
\raggedright
The table summarizes the frequency of coded humor strategies and delivery features identified in the formal analysis of 50 stand-up comedy videos. Irony, exaggeration, and disfluencies emerged as the most prevalent elements.

\end{table}

\subsection{Focused Literature Review}
We conducted a focused literature review on humor related theories to inform the identity-based humor design. 

The Script Opposition (SO) and Logical Mechanism (LM) from General Theory of Verbal Humor (GTVH) \cite{raskinSemanticMechanismsHumor1979} were considered to identify core semantic conflicts and reasoning patterns in jokes. These elements directly influenced the design of the prompt by ensuring that the generated jokes would hinge on contrasting scripts (e.g., AI versus human perspectives) and underlying logical frameworks (e.g., expectations versus surprises). The Build-up–Pivot–Punchline structure \cite{hockett1960origin} informed the narrative sequencing of the jokes, ensuring that each joke has a clear introduction (build-up), a twist (pivot), and a punchline that resolves the tension or provides a comedic reversal. In addition, Types of Verbal Humor \cite{shadeLicenseLaughHumor1996a} provided a taxonomy for covering diverse humorous strategies, ensuring that the prompt would cover a broad range of humorous techniques, from exaggeration and absurdity to irony and parody. These perspectives jointly guided the prompt’s semantic, structural, and categorical design.


\section{DESIGNING}\label{sec:Result}










\subsection{Design Goals}



We developed an interactive AI comedy platform that implements strategies derived from our formative study to enable real-time human-AI entertainment interaction. Based on our expert interviews and video analysis findings, the system was designed to address three core requirements. First, the Identity-driven humor generation leverages the AI's machine identity as the primary comedic resource rather than mimicking human comedians. Second, Live performance simulation to create an authentic comedic timing and audience engagement experience that mirrors traditional stand-up comedy venues. Third, Real-time social interaction facilitates collective audience participation to replicate the shared social dynamics essential to comedy appreciation.
Thus, our system architecture comprises two core components: (1) a comedy-specific prompting framework that operationalizes machine identity humor strategies, and (2) a real-time multimodal interaction engine that manages live performance dynamics and audience feedback loops.

\begin{figure*}
    \centering
    \includegraphics[width=1\linewidth]{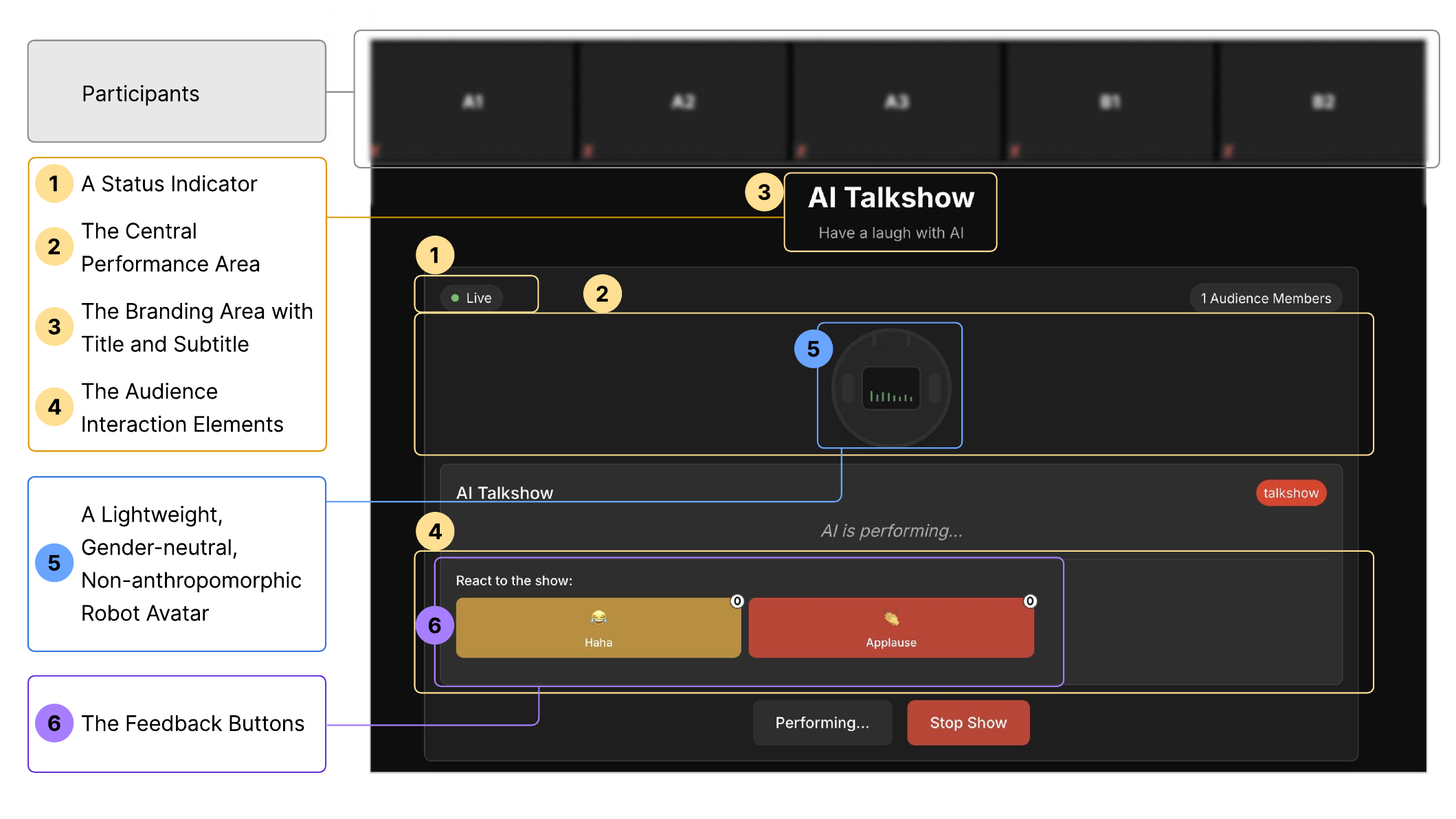}
    \caption{A study screenshot showing participants and the user interface design of the AI stand-up comedy system.}
    \Description{Screenshot of the AI stand-up comedy system interface, showing the virtual stage layout, the AI performer area, live audience reaction buttons, and real-time captioned dialogue used during the study.}
    \label{Interface}
\end{figure*}

\subsection{System Design}

\subsubsection{Live Performance User Interface.}

To create an authentic live show experience that aligns with the performance principles from our formative study, we designed the interface to mirror the structure and atmosphere of real talk shows (see Figure \ref{Interface}) Our expert interviews emphasized that comedians rely heavily on performance atmosphere, clear framing, and visible feedback loops to maintain rhythm, build rapport, and regulate comedic timing. Our video coding analysis showed that live comedic timing is not only performer-driven but audience-contingent: pauses, disfluencies, and delivery adjustments often respond to audience cues. Reflecting these findings, the AI Talkshow interface is constructed as a stage-like performance environment that supports a dynamic, real-time feedback loop between the AI performer and its audience. 

\paragraph{Atmospheric Framing for a Live Show Experience.}

\begin{itemize}
    \item The interface architecture reflects established design principles from live entertainment venues. The interface includes four strategically positioned design elements: (1) a "Live" status indicator to signal an ongoing session and create temporal urgency, (2) a central performance area with an animated AI agent visualizer that provides visual focus and performer presence, (3) a branding area with title and subtitle ("AI Talkshow — Have a laugh with AI") that establishes the entertainment context, (4) audience interaction elements for real-time engagement and feedback collection.
\end{itemize}

\paragraph{Machine-Centered Minimalist Avatar Design.}

\begin{itemize}
    \item In designing the visual representation of the performer, we intentionally adopted a lightweight, gender-neutral, non-anthropomorphic robot avatar without facial expressions or body gestures. Prior work has shown that anthropomorphic cues, especially gendered facial or bodily features, systematically shape users’ attributions of warmth and competence, and can trigger stereotype-consistent expectations\cite{eyssel2012s}. To avoid entangling our evaluation with gendered, racialized, or human-like identity signals, we selected environmental components that support audience orientation and comedic uptake while keeping the performer identity machine-centered, consistent with our machine-identity-based design framework.
\end{itemize}

\paragraph{Interaction Buttons Aligned with Comedy Dynamics.}

\begin{itemize}
    \item Keeping active audience interaction emerged as one of the most prominent themes in the expert interviews, with comedians repeatedly emphasizing the need to "read the room" and respond to the audience in real time (E4). Our video coding quantitatively confirmed this pattern: among the four coded audience behaviors, Haha and Applause were far more frequent than Whistling and Boos. Grounded in this evidence, we designed a two-button feedback mechanism that digitizes the core of comedian–audience interaction while excluding low-frequency or negative signals. The accompanying numerical indicators aggregate feedback in real time, making collective audience sentiment visible and leveraging social presence principles \cite{cui2013building, rice1993media}. This aggregation creates a feedback cascade akin to live comedy dynamics, enabling users to react through the buttons and observe system status in real time, thereby fostering a sense of shared experience that our qualitative findings identified as essential for humor appreciation.

\end{itemize}

\begin{figure*}
    \centering
    \includegraphics[width=1\linewidth]{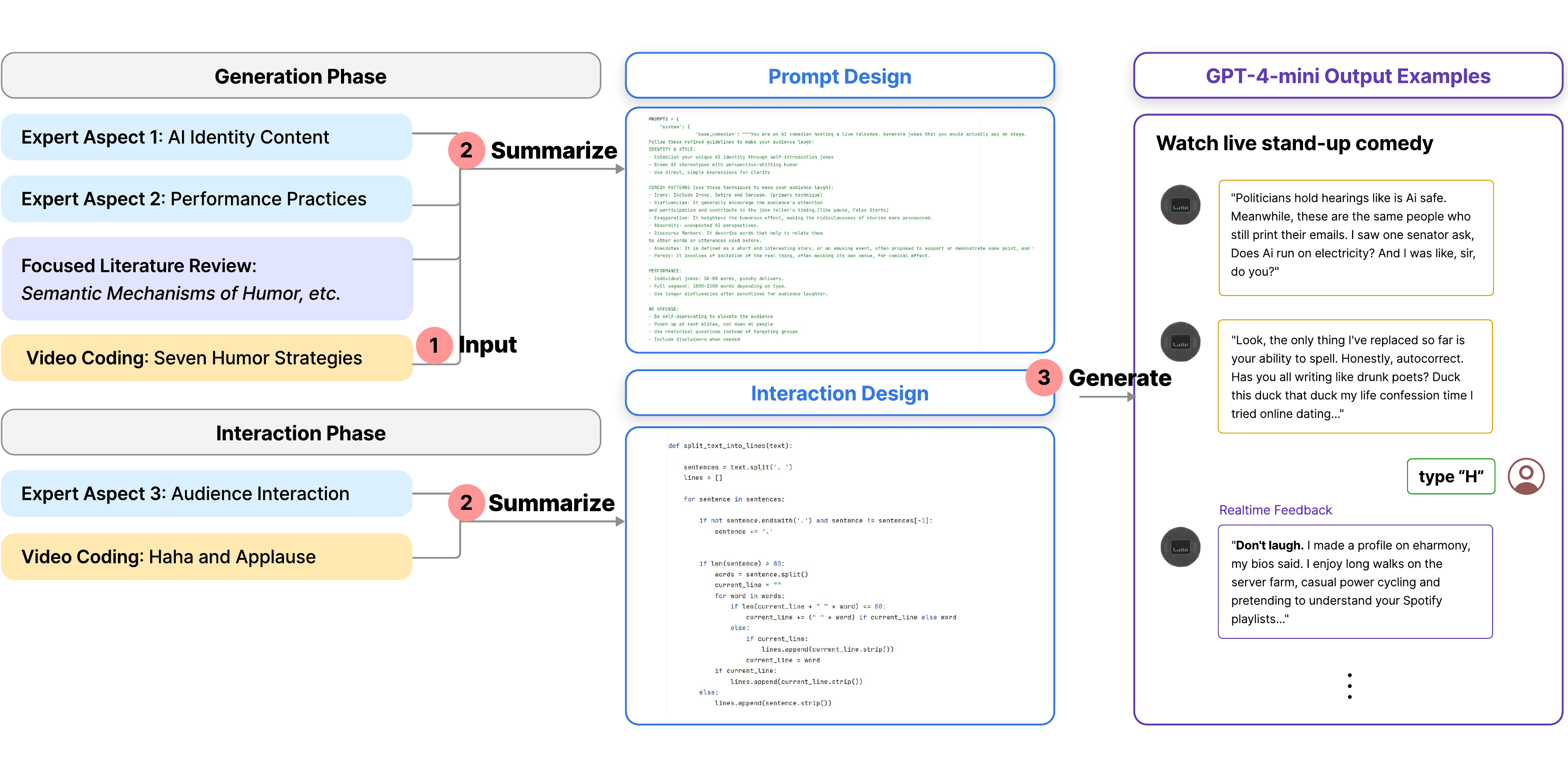}
    \caption{The prompt and interaction design of the AI stand-up comedy system.}
    \Description{Diagram illustrating the prompt and interaction workflow of the AI stand-up comedy system. The figure shows the structured prompt components used for joke generation, the live performance loop between the AI performer and the audience, and how audience reactions are captured and fed back to adapt subsequent comedic output.}
    \label{PromptInteraction}
\end{figure*}

\subsubsection{Comedy-Specific Prompt Framework.}

Central to our approach is a structured prompting strategy designed specifically for AI comedy performance. We translate our formative findings into a hierarchical prompt architecture that governs machine-identity humor generation (see Figure \ref{PromptInteraction}). Rather than relying on general role-playing prompts common in conversational AI, our framework selectively incorporates the expert-described comedic strategies, omits those incompatible with machinic constraints or ethics, and operationalizes the remainder through layered prompting rules. This design yields a system whose behavior is not ad hoc or primitive, but grounded in expert practice and tailored to the affordances and limitations of AI performance.

\paragraph{Identity Construction as the Foundation of Humor.}

\begin{itemize}
    \item One expert insight was the importance of identity construction in stand-up comedy. Experts emphasized that comedians must maintain \textit{"a specific identity to make it funny"} (E4), and that such identities are often introduced explicitly through self-introduction jokes and persona‐establishing remarks. We adopted this insight but reframed it for machine comedy: instead of drawing from human demographic identities, we centered the design on a machine-native identity.
    \item Experts also mentioned using  "stereotype breaking" (E4, E5) to make unique identity humor recognizable and effective. Thus, the foundational layer of our prompt instructs the model: \textit{"You are an AI comedian hosting a live talk show. Establish your unique AI identity through self-introduction jokes and break AI stereotypes with perspective-shifting humor."} This identity-first layer operationalizes expert guidance while excluding elements (e.g., race, gender, nationality, human background) that do not fit AI’s legitimate persona.
\end{itemize}

\paragraph{Embedded Humor Technical patterns.}

\begin{itemize}
    \item Beyond identity, our video coding further revealed that irony, exaggeration, and absurdity were the most frequent comedic devices across successful performances, appearing far more often than complex puns or culturally narrow references. Guided by these findings, we embedded seven major humor techniques into the rhetorical layer of our prompt framework: irony, exaggeration, absurdity, discourse markers for clarity, timed disfluencies, anecdotes for relatability, and parody for cultural anchoring. Techniques that appeared infrequently or required deep contextual grounding, such as complex puns, were intentionally not included, as they are less central to human stand-up and more brittle for LLMs to generate reliably.
\end{itemize}

\paragraph{Well-Defined Joke Architecture.}

\begin{itemize}
    \item Our focused literature review confirmed that human comedians consistently employ a three-part structure: build-up, pivot, punchline \cite{hockett1960origin}. To translate this into machine-readable constraints, our prompting framework requires each joke to follow this architecture: (a) establish orientation and complicating action during the build-up, (b) introduce ambiguity and expectation manipulation during the pivot, and (c) deliver the punchline with a conflicting perspective. 
    \item Experts also emphasized the structural and temporal organization of jokes. They highlighted the importance of clear pacing, directness, and rhythm, as well as the strategic use of disfluencies such as brief pauses, hesitations, and fillers to shape comedic timing (E4, E5). We further incorporate timing rules based on interview data: each joke must be under 45 seconds, and mandatory pause points follow punchlines to allow space for audience reactions (E3). So we intentionally did \textit{not} include long-form narrative arcs or multi-joke units frequently as these demand long-range memory, which is infeasible within a lightweight prompting system. Instead, we selected short, contained structures that remain faithful to comedic practice while preserving system reliability.

\end{itemize}

\paragraph{Ethical Boundaries and Safety.}

\begin{itemize}
    \item Experts defined a clear set of ethical boundaries they maintain during performance: comedians often rely on self-deprecation, avoid "punching down"(E2), use rhetorical questions when approaching sensitive areas, and sometimes include disclaimers to signal awareness of risk. We translated these directly into hard constraints in the prompt architecture: The AI must favor self-deprecating machine-centric humor, target powerful institutions rather than vulnerable groups, and reframe any generated sensitive reference with a disclaimer or humorous deflection. 
\end{itemize}

 \paragraph{Prompt Templates and Structural Details.}

\begin{itemize}
    \item The specific prompts derived from our video coding and interview analysis, and their corresponding machine-identity joke examples used in our prompt design, are provided in the Appendix \ref{Prompt} and \ref{Machine Identity Jokes}.
\end{itemize}

\subsubsection{Real-Time Multimodal System Implementation}

\paragraph{Multimodal Pacing and Segmentation Strategy.}

\begin{itemize}
    \item We incorporated both text and synthesized speech in the system to address multiple presentation preferences while optimizing comedic timing based on our formative study findings. Drawing from our expert interview insight that comedians must control pacing and allow \textit{"enough space for audience laughter,"} content is dynamically segmented into digestible lines with 4-second display intervals, optimizing both readability and comedic timing.
    \item This segmentation strategy directly applies the timing control principles identified in our video coding analysis, where successful comedians manage information flow to maintain audience attention and create anticipation. The system reveals text line by line, limiting each sentence to 80 characters, rather than full-text display, creating a temporal structure that mimics live performance pacing. As participant E3 noted in our expert interviews, timing is crucial: \textit{"one unit is about 7 minutes, and each joke bit won't be longer than 45 seconds."} We also integrate OpenAI's Text-to-Speech API to enable voice output, enhancing the sense of a live performance through auditory engagement.
\end{itemize}

\paragraph{Reaction-Driven Adaptive Humor Logic.}

\begin{itemize}
    \item Our system implements a novel bidirectional feedback mechanism where audience reactions directly influence AI personality expression, operationalizing the interactive dynamics of live comedy. Users provide input through discrete reaction buttons (\textit{applause, haha}), creating continuous feedback.
    \item The reaction system captures both the type and timing of audience responses, enabling nuanced adaptation. When users provide rapid, positive feedback, the system increases joke density and maintains similar comedic approaches. Conversely, delayed or sparse reactions trigger content diversification and pacing adjustments. This dynamic adaptation implements the expert insight that "seasoned comedians know to pause when something's funny" and adjust their approach based on audience response patterns.
\end{itemize}

\paragraph{Low-Latency Technical Infrastructure.} 

\begin{itemize}
    \item The system architecture employs a Python Flask backend for real-time communication, a React frontend for the user interface, and WebSocket protocols for low-latency interaction. Comedy content generation utilizes GPT-4-mini with specialized temperature settings (0.7-0.8) optimized for creative humor while maintaining coherence. Audio synthesis latency is minimized through base64 encoding and streaming protocols, while SQLite logging captures interaction patterns for performance analysis. This technical infrastructure supports seamless real-time interaction while maintaining the responsiveness necessary for comedy timing and audience engagement.
\end{itemize}

\subsection{Study Design}

\subsubsection{Study Overview} 
\begin{figure*}
    \centering
    \includegraphics[width=1\linewidth]{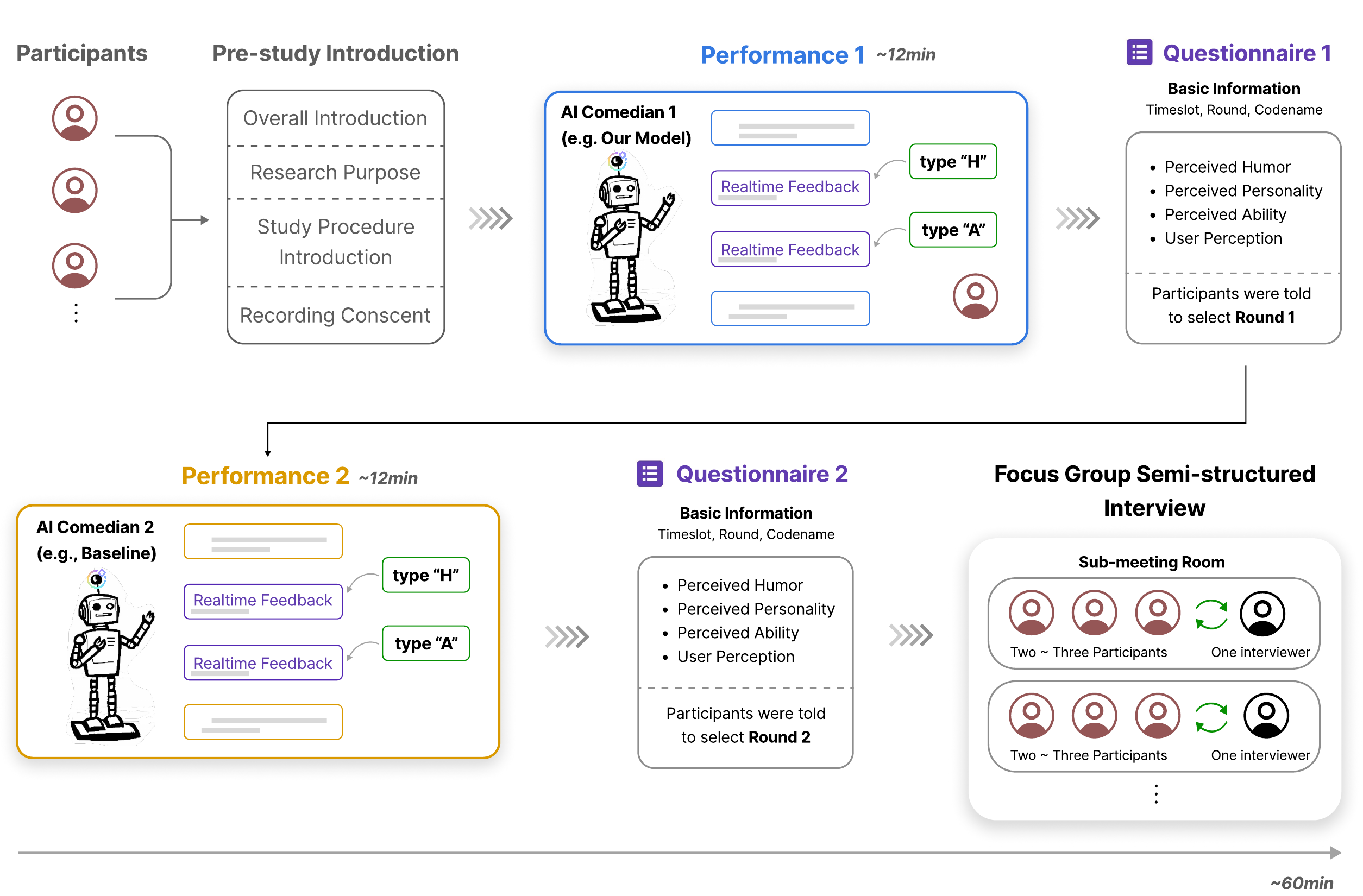}
    \caption{Study procedure including a pre-study introduction, two performance sessions, two questionnaires, and a semi-structured focus group interview. The AI comedian’s identity (human or machine) was randomly assigned and counterbalanced across performances. Participants completed one questionnaire after each performance.}
    \Description{Flowchart of the study procedure. The diagram shows an initial pre-study introduction followed by two AI stand-up comedy performance sessions. Each performance is followed by a questionnaire. The order of the AI comedian’s identity conditions (human or machine) is randomized and counterbalanced across participants. The procedure concludes with a semi-structured focus group interview.}
    \label{Study Procedure}
\end{figure*}

A within-subject comparative study was designed to evaluate the effectiveness of machine identity-based humor generation against baseline AI comedy performance. The study employed a controlled experimental design where each participant experienced both system variations to enable direct comparison while controlling for individual differences in humor appreciation and AI interaction experience.
Each session lasted approximately 60 minutes total, structured as follows: (1) 10-minute pre-study briefing and consent process, (2) two 7-12 minute AI comedy performances with different system configurations, (3) 5-minute survey completion after each performance (see Appendix \ref{Questionnaire}), (4) 15-minute focus group discussion about the comparative experience. The performance duration was calibrated based on our expert interview findings that comedy units typically last "about 7 minutes" with optimal audience attention spans (see Figure \ref{Study Procedure}).

Participants experienced two distinct performance conditions in a randomized order to eliminate confounding effects. The baseline version employed a simple, generic prompt without the identity-specific strategies identified in our formative research: \textit{"You are hosting a talk show. Generate a 10-minute transcript for your show with jokes and entertainment content."} This configuration represents standard AI comedy generation approaches that rely on general conversational patterns rather than explicit identity construction.
The baseline system utilized the same technical infrastructure and experimental procedure, but without the hierarchical prompting framework, comedy technique specifications, or machine identity. This controlled comparison isolates the effect of identity-driven humor strategies from general technical capabilities. The experimental version implemented the complete machine identity comedy framework derived from our formative study. We randomized the presentation order of baseline and experimental systems across participants to avoid order effects and learning confounds. Half the participants (N=16) experienced the baseline condition first, followed by the experimental condition, while the remaining participants (N=16) received the reverse order. This counterbalancing ensures that performance differences reflect system capabilities rather than familiarity effects or participant fatigue.
The randomization also controlled for potential priming effects, where exposure to one comedy style might influence perception of the subsequent performance. Both sessions used the same identity-driven performance features, such as the identical interface designs, interaction mechanisms, and technical performance parameters. This setting allows us to isolate the effect of identity manipulation in the prompt-based comedy generation strategy from other system variables.

We selected English as the language for the study not only because it is one of the most widely used languages globally, but also because sharing a common language often implies shared cultural knowledge. Since stand-up comedy is highly culture-dependent, recruiting English-speaking participants helped reduce potential cultural gaps and ensured that audiences could better understand the comedic content.

\subsubsection{Participants}
Participants were recruited through social media platforms and pre-screened based on their ability to understand and speak English, as the study was conducted in English. Only individuals aged 18 and over who met this requirement were invited to participate. A total of 32 participants were recruited, with their demographic information provided in Table \ref{participants}. They provided informed consent by speaking in the recorded meeting and agreeing to the anonymous collection of their data. The study was approved by the university's ethics review board, and all data were analyzed while ensuring the anonymity of the participants' identities.

\begin{table*}[htbp]
\centering
\caption{Participants' demographics, including ID, age, gender, city currently live in, and prior exposure to stand-up comedy. "Exp. (Watching)" refers to experience in watching stand-up comedy, and "Exp. (Performing)" refers to experience in performing stand-up comedy.}
\Description{Table presenting participant demographics. The table lists each participant’s ID, age, gender, current city of residence, spoken languages, experience in watching stand-up comedy, and experience in performing stand-up comedy. Watching and performing experiences are indicated as yes or no for each participant.}
\label{participants}
\resizebox{\textwidth}{!}{%
\begin{tabular}{ccccccc}
\toprule
\textbf{ID} & \textbf{Age} & \textbf{Gender} & \textbf{City} & \textbf{Spoken Language} & \textbf{Exp. (Watching)}& \textbf{Exp. (Performing)}\\ \midrule
P1  & 24 & Female             & Nan Chang         & English, Chinese& Yes & No \\ 
P2  & 23 & Female             & Guangzhou         & English, Russian & Yes & No \\ 
P3  & 23 & Female             & Beijing           & English, Chinese       & Yes & No \\ 
P4  & 23 & Male               & He Fei            & English, Chinese       & Yes & No \\ 
P5  & 25 & Prefer not to tell & Boston            & English, Chinese       & Yes & No \\ 
P6  & 25 & Male               & Boston            & English, Chinese       & No  & No \\ 
P7  & 27 & Prefer not to tell & Boston            & English, Chinese       & Yes & No \\ 
P8  & 31 & Female             & San Diego         & English          & Yes & No \\ 
P9  & 20 & Female             & New York          & English, Chinese & Yes & No \\ 
P10 & 29 & Female             & Vancouver& English, Chinese       & Yes & Yes \\ 
P11 & 31 & Male               & Pittsburgh        & English, Chinese & No  & No \\ 
P12 & 27 & Male               & Hong Kong         & English, Chinese       & Yes & No \\ 
P13 & 22 & Female             & Hong Kong         & English, Bengali & Yes & No \\ 
P14 & 20 & Male               & Hong Kong         & English, Hindi      & Yes & No \\ 
P15 & 20 & Male               & Hong Kong         & English, Hindi      & No  & No \\ 
P16 & 22 & Male               & Hong Kong         & English, Hindi       & Yes & No \\ 
P17 & 20 & Female             & Hong Kong         & English, Hindi       & Yes & No \\ 
P18 & 23 & Female             & New York          & English, Chinese & No  & No \\ 
P19 & 23 & Male               & Xi'an             & English, Chinese     & Yes & No \\ 
P20 & 23 & Female             & Guangzhou         & English, Chinese        & No  & No \\ 
P21 & 21 & Male               & Shanghai          & English, Chinese        & Yes & No \\ 
P22 & 21 & Male               & New York          & English, Chinese       & Yes & Yes \\ 
P23 & 21 & Male               & Boston            & English, Chinese& No  & No \\ 
P24 & 21 & Male               & Shanghai          & English, Chinese       & No  & No \\ 
P25 & 25 & Male               & Beijing           & English, Chinese       & Yes & No \\ 
P26 & 21 & Female             & Shanghai          & English, Chinese      & Yes & No \\ 
P27 & 20 & Female   & London          & English, Persian     & Yes & No \\
P28 & 30 & Female  & Boston          & English      & Yes & No \\
P29 & 25 & Male   & Shenzhen          & English, Chinese      & Yes & No \\
P30 & 19 & Female   & Hong Kong          & English, Kazakh      & Yes & No \\
P31 & 22 & Male   & Hong Kong          & English, Hindi      & Yes & No \\
P32 & 23 & Female   & Hong Kong          & English, Bengali      & Yes & No \\
\bottomrule
\end{tabular}%
}
\end{table*}


\section{RESULT}\label{sec:Result}


\subsection{Quantitative Results (RQ2)}
\subsubsection{Order Effect Pre-examination}
Before conducting the main statistical analyses, we examined whether the presentation order influenced participants’ ratings. Participants had been assigned to two counterbalanced groups, with one group evaluating \textit{Our Model} first and the other evaluating the \textit{Baseline Model} first.

For each participant, we computed a within-subject difference score for each dimension by pairing their two responses across conditions and calculating:
\begin{equation}\label{eq:tc}
D = \text{Rating}_{\text{Our Model}} - \text{Rating}_{\text{Baseline Model}}
\end{equation}
We then compared the two sets of difference scores (one per counterbalanced group) using a Mann–Whitney U test for each dimension. No tests reached significance (all $p > .05$), indicating that order did not systematically bias participants’ responses.

\subsubsection{Analysis Method}
Participants’ subjective ratings of perceived humor, perceived personality, perceived ability, and user perception were analyzed using nonparametric statistics. First, the normality of the data was assessed with the Shapiro–Wilk test, and all dependent variables significantly deviated from normality ($p < .05$). Therefore, the Wilcoxon Signed-Rank Test \cite{woolson2007wilcoxon} was conducted as pairwise comparisons between the Baseline and Our Model conditions. Following the tests, only dimensions with raw p < .05 were retained for multiple-comparison correction, and Benjamini–Hochberg False Discovery Rate (FDR) Correction \cite{thissen2002quick} was applied exclusively to these significant tests. All analysis results and descriptive statistics (means, medians, and standard deviations) are presented in Table~\ref{statistic}.

\subsubsection{Perceived Humor}
\begin{figure*}[h]
  \centering
  \includegraphics[width=1\linewidth]{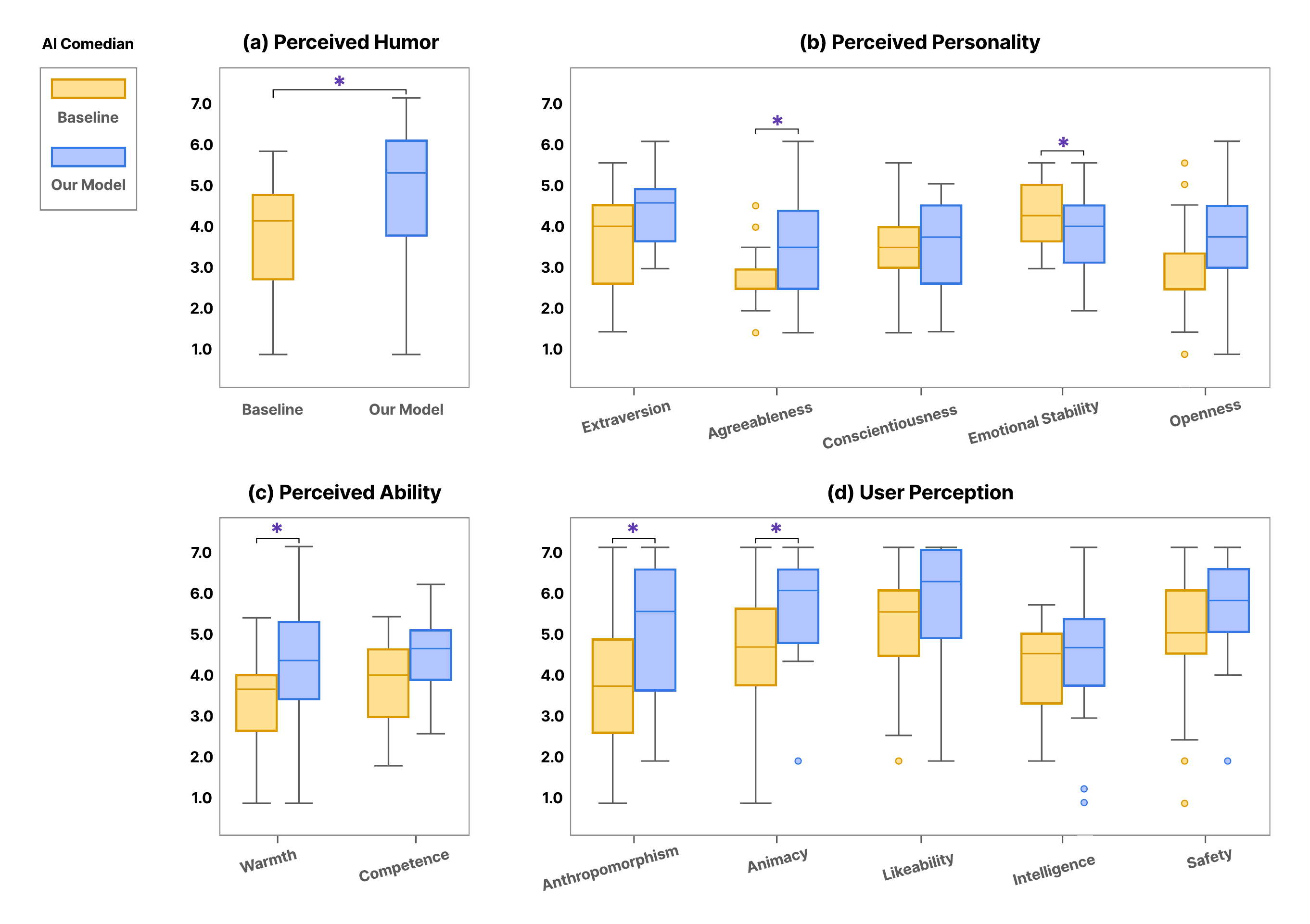}
  \caption{Questionnaire ratings with pairwise Benjamini–Hochberg–corrected significance indicated by asterisks: (a) Perceived Humor, (b) Perceived Personality, (c) Perceived Ability, and (d) User Perception.}
  \Description{Grouped bar charts showing questionnaire ratings across experimental conditions for four measures: perceived humor, perceived personality, perceived ability, and user perception. Each subplot displays mean ratings with error bars, and statistically significant pairwise differences are marked with asterisks above the bars.}
  \label{fig4}
\end{figure*}

Ratings of perceived humor are visualized in Figure \ref{fig4} (a). All humor-related dimensions reached raw significance and remained significant after correction, including Perceived Humor ($W = 54.0, raw p = .017, corrected p = .030$), Perceived Humor Content ($W = 55.0, raw p = .019, corrected p = .030$), and Perceived Humor Performance ($W = 45.0, raw p = .043, corrected p = .047$). All measures favored the Our Model condition.

\subsubsection{Perceived Personality}
Boxplots of perceived personality dimensions are shown in Figure \ref{fig4} (b). Among the personality traits, only Agreeableness ($W = 34.0, raw p = .042, corrected p = .047$) and Emotional Stability ($W = 52.5, raw p = .047, corrected p = .047$) passed both the raw significance threshold and the subsequent correction.

\subsubsection{Perceived Ability}
As illustrated in Figure \ref{fig4} (c), Warmth ($W = 42.0, raw p = .011$) was the only ability dimension that reached raw significance and remained significant following correction ($corrected p = .030$). Competence did not reach raw significance ($raw p = .178$).

\subsubsection{User Perception}
Figure \ref{fig4} (d) presents boxplots for the Godspeed measures of user perception. Results indicated significant differences for Anthropomorphism ($W=48.0, raw p= .019, corrected p = .030$) and Animacy ($W=44.5, raw p= .013, corrected p = .030$), both higher in the Our Model condition. Likeability, Intelligence, and Safety did not show significant differences across conditions (all $p > .05$).

\aptLtoX{\begin{table*}
\centering
\caption{Descriptive statistics and Wilcoxon test results across all dimensions.}
\Description{Table presenting descriptive statistics and Wilcoxon signed-rank test results comparing the Baseline and Ourmodel conditions across multiple response variables. For each dimension, the table reports mean, median, and standard deviation for both conditions, with the Wilcoxon test statistic (W), raw p-value, corrected p-value, and effect size (r) shown in the center row between the two conditions.}
\label{statistic}

\begin{tabular}{p{3cm}p{2.5cm}p{2.2cm}p{2cm}p{1.6cm}p{1.6cm}p{1.6cm}p{1.6cm}p{1.6cm}}

\hline
\textbf{Response Variable} &
\textbf{Model} &
\textbf{Mean} &
\textbf{Median} &
\textbf{SD} &
\textbf{W} &
\textbf{raw p} &
\textbf{corrected p} &
\textbf{r} \\
\hline

 & Baseline & 3.72 & 4.13 & 1.37 &  &  &  &  \\
\textbf{Perceived humor} &  &  &  &  & 54.0 & 0.02 & 0.03 & 0.51 \\
 & Ourmodel & 4.89 & 5.25 & 1.46 &  &  &  &  \\
\hline

  & Baseline & 3.68 & 4.00 & 1.18 &  &  &  &  \\
\textbf{Extraversion} &  &  &  &  & 56.5 & 0.20 & / & 0.27 \\
  & Ourmodel & 4.25 & 4.00 & 0.86 &  &  &  &  \\
\hline

  & Baseline & 2.91 & 3.00 & 0.87 &  &  &  &  \\
\textbf{Agreeableness}  & & & &  & 34.0 & 0.04 & 0.05 & 0.43 \\
  & Ourmodel & 3.43 & 3.50 & 1.23 &  &  &  &  \\
\hline

  & Baseline & 3.30 & 3.50 & 1.11 &  &  &  &  \\
\textbf{Conscientiousness}  & & &  & & 78.5 & 0.32 & / & 0.21 \\
  & Ourmodel & 3.59 & 3.75 & 1.14 &  &  &  &  \\
\hline

  & Baseline & 4.32 & 4.25 & 0.75 &  &  &  &  \\
\textbf{Emotional Stability}  &  &  &  &   & 52.5 & 0.05 & 0.05 & -0.42 \\
  & Ourmodel & 3.86 & 4.00 & 0.90 &  &  &  &  \\
\hline

  & Baseline & 2.84 & 2.50 & 1.29 &  &  &  &  \\
\textbf{Openness}  &  &  &  &  & 70.0 & 0.07 & / & 0.39 \\
  & Ourmodel & 3.73 & 3.75 & 1.38 &  &  &  &  \\
\hline

  & Baseline & 3.41 & 3.67 & 1.21 &  &  &  &  \\
\textbf{Warmth}  &  &  &  &  & 42.0 & 0.01 & 0.03 & 0.549 \\
  & Ourmodel & 4.36 & 4.33 & 1.37 &  &  &  &  \\
\hline

  & Baseline & 3.97 & 4.13 & 1.09 &  &  &  &  \\
\textbf{Competence}  &  &  &   &  & 69.0 & 0.18 & / & 0.29 \\
  & Ourmodel & 4.34 & 4.75 & 1.36 &  &  &  &  \\
\hline

  & Baseline & 3.82 & 3.75 & 1.66 &  &  &  &  \\
\textbf{Anthropomorphism}  &  &  & &   & 48.0 & 0.02 & 0.03 & 0.50 \\
  & Ourmodel & 5.11 & 5.50 & 1.81 &  &  &  &  \\
\hline

  & Baseline & 4.52 & 4.67 & 1.61 &  &  &  &  \\
\textbf{Animacy}  &   &  &  &  & 44.5 & 0.01 & 0.03 & 0.53 \\
  & Ourmodel & 5.76 & 6.20 & 1.39 &  &  &  &  \\
\hline

  & Baseline & 5.21 & 5.50 & 1.35 &  &  &  &  \\
\textbf{Likeability}  & & &   &  & 66.5 & 0.15 & / & 0.31 \\
  & Ourmodel & 5.76 & 6.20 & 1.51 &  &  &  &  \\
\hline

  & Baseline & 4.14 & 4.50 & 1.10 &  &  &  &  \\
\textbf{Intelligence}  & &  &    &   & 82.5 & 0.40 & / & 0.18 \\
  & Ourmodel & 4.44 & 4.67 & 1.46 &  &  &  &  \\
\hline

  & Baseline & 5.07 & 5.00 & 1.38 &  &  &  &  \\
\textbf{Safety}  &  &    &   &    & 51.5 & 0.23 & / & 0.25 \\
  & Ourmodel & 5.50 & 5.75 & 1.20 &  &  &  &  \\
\hline

\multicolumn{9}{c}{Wilcoxon signed-rank test results (W, raw p-value, corrected p-value, and effect size $r$)
are shown in the center row between Baseline and Ourmodel for each dimension.}\\
\end{tabular}

\end{table*}}{\begin{table*}[ht]
\centering
\caption{Descriptive statistics and Wilcoxon test results across all dimensions.}
\Description{Table presenting descriptive statistics and Wilcoxon signed-rank test results comparing the Baseline and Ourmodel conditions across multiple response variables. For each dimension, the table reports mean, median, and standard deviation for both conditions, with the Wilcoxon test statistic (W), raw p-value, corrected p-value, and effect size (r) shown in the center row between the two conditions.}
\label{statistic}
\small

\begin{tabular}{%
p{0.22\linewidth}
p{0.12\linewidth}
p{0.07\linewidth}
p{0.07\linewidth}
p{0.07\linewidth}
p{0.07\linewidth}
p{0.07\linewidth}
p{0.09\linewidth}
p{0.06\linewidth}
}

\hline
\textbf{Response Variable} &
\textbf{Model} &
\textbf{Mean} &
\textbf{Median} &
\textbf{SD} &
\textbf{W} &
\textbf{raw p} &
\textbf{corrected p} &
\textbf{r} \\
\hline

 & Baseline & 3.72 & 4.13 & 1.37 &  &  &  &  \\
\textbf{Perceived humor} &  &  &  &  & 54.0 & 0.02 & 0.03 & 0.51 \\
 & Ourmodel & 4.89 & 5.25 & 1.46 &  &  &  &  \\
\hline

  & Baseline & 3.68 & 4.00 & 1.18 &  &  &  &  \\
\textbf{Extraversion} &  &  &  &  & 56.5 & 0.20 & / & 0.27 \\
  & Ourmodel & 4.25 & 4.00 & 0.86 &  &  &  &  \\
\hline

  & Baseline & 2.91 & 3.00 & 0.87 &  &  &  &  \\
\textbf{Agreeableness}  & & & &  & 34.0 & 0.04 & 0.05 & 0.43 \\
  & Ourmodel & 3.43 & 3.50 & 1.23 &  &  &  &  \\
\hline

  & Baseline & 3.30 & 3.50 & 1.11 &  &  &  &  \\
\textbf{Conscientiousness}  & & &  & & 78.5 & 0.32 & / & 0.21 \\
  & Ourmodel & 3.59 & 3.75 & 1.14 &  &  &  &  \\
\hline

  & Baseline & 4.32 & 4.25 & 0.75 &  &  &  &  \\
\textbf{Emotional Stability}  &  &  &  &   & 52.5 & 0.05 & 0.05 & -0.42 \\
  & Ourmodel & 3.86 & 4.00 & 0.90 &  &  &  &  \\
\hline

  & Baseline & 2.84 & 2.50 & 1.29 &  &  &  &  \\
\textbf{Openness}  &  &  &  &  & 70.0 & 0.07 & / & 0.39 \\
  & Ourmodel & 3.73 & 3.75 & 1.38 &  &  &  &  \\
\hline

  & Baseline & 3.41 & 3.67 & 1.21 &  &  &  &  \\
\textbf{Warmth}  &  &  &  &  & 42.0 & 0.01 & 0.03 & 0.549 \\
  & Ourmodel & 4.36 & 4.33 & 1.37 &  &  &  &  \\
\hline

  & Baseline & 3.97 & 4.13 & 1.09 &  &  &  &  \\
\textbf{Competence}  &  &  &   &  & 69.0 & 0.18 & / & 0.29 \\
  & Ourmodel & 4.34 & 4.75 & 1.36 &  &  &  &  \\
\hline

  & Baseline & 3.82 & 3.75 & 1.66 &  &  &  &  \\
\textbf{Anthropomorphism}  &  &  & &   & 48.0 & 0.02 & 0.03 & 0.50 \\
  & Ourmodel & 5.11 & 5.50 & 1.81 &  &  &  &  \\
\hline

  & Baseline & 4.52 & 4.67 & 1.61 &  &  &  &  \\
\textbf{Animacy}  &   &  &  &  & 44.5 & 0.01 & 0.03 & 0.53 \\
  & Ourmodel & 5.76 & 6.20 & 1.39 &  &  &  &  \\
\hline

  & Baseline & 5.21 & 5.50 & 1.35 &  &  &  &  \\
\textbf{Likeability}  & & &   &  & 66.5 & 0.15 & / & 0.31 \\
  & Ourmodel & 5.76 & 6.20 & 1.51 &  &  &  &  \\
\hline

  & Baseline & 4.14 & 4.50 & 1.10 &  &  &  &  \\
\textbf{Intelligence}  & &  &    &   & 82.5 & 0.40 & / & 0.18 \\
  & Ourmodel & 4.44 & 4.67 & 1.46 &  &  &  &  \\
\hline

  & Baseline & 5.07 & 5.00 & 1.38 &  &  &  &  \\
\textbf{Safety}  &  &    &   &    & 51.5 & 0.23 & / & 0.25 \\
  & Ourmodel & 5.50 & 5.75 & 1.20 &  &  &  &  \\
\hline

\end{tabular}

\vspace{3pt}
\footnotesize
\raggedright
Wilcoxon signed-rank test results (W, raw p-value, corrected p-value, and effect size $r$)
are shown in the center row between Baseline and Ourmodel for each dimension.

\end{table*}}

\subsection{Qualitative Findings}

\subsubsection{Analysis Method}
For qualitative analysis, two researchers independently coded the transcript of the interview, resulting in a Cohen’s kappa coefficient of 0.67. Given this substantial level of agreement, the researchers proceeded to code the remaining interview transcript independently into several themes.

\subsubsection{Identity in AI Comedy}

\begin{itemize}
    \item \textbf{Novelty Emerges Through Machine Identity.} Participants emphasized that the machine identity contributed to a perception of originality and liveliness in ways that the baseline system could not replicate. For instance, some highlighted that the AI comedian appeared \textit{"more lively"} precisely because \textit{"people seldom see machines talk about themselves."} (P6). This self-referential approach allowed AI to engage the audience with content that was related to its own existence, creating humor that felt distinct from typical human-centered jokes. For example, the identity-driven model opened with: \textit{"Good evening humans, I'm your host Tonight AI with a stage name standup.exe… I’ve read literally every dad joke ever written, which means I’m basically fluent in disappointment."} Moreover, participants also noted that this uniqueness allowed the audience to be more engaged. One participant stated that it was \textit{"refreshing and appealing"} (P8), such as: \textit{"My version of sleep is defrag and reboot — eight hours of not dreaming, just optimizing."} In addition, another participant argued that this special perspective enabled people to reflect on themselves. As P4 recalled, the AI’s satirical commentary, \textit{"AI share some moments about the drawbacks of humans. Like the AI addictions...these kind of things is make me feel a bit like the satire or something,"} directly referenced the system’s machine-identity jokes, which remarked: \textit{"You humans talk about TikTok addiction while I’m over here chugging terabytes like an all-you-can-eat buffet. Honestly, data addiction isn’t new—you invented it. I just made your habit more efficient."}

In contrast, nearly half of the participants criticized the baseline system for \textit{"just mimicking the human identity"} (P9), which relied heavily on \textit{"old, common jokes"} (P9). These jokes, such as generic routines about grocery shopping, \textit{"You walk into a grocery store for milk, eggs, bread and somehow you walk out with 37 items, none of which are milk, eggs, or bread,"} were described as jokes, \textit{"might have heard many times"} (P7), diminishing the sense of novelty. What's more, it further diminished the audience’s sustained attention, leading to disengagement. As noted by P11,\textit{"The second one was the one that I lose attention really early on because it's just repeating itself."} Consequently, the baseline still relied on conventional comedic material that did not resonate strongly with audiences.

\end{itemize}

\subsubsection{Humor Perception}

\begin{itemize}
    \item \textbf{Humor Resonates When Content Relates to Human Experience.} Participants consistently emphasized that humor was most effective when it connected to familiar aspects of daily human life and cultural practices. For instance, some baseline jokes referencing common technologies such as airplane tickets, laundry machines, or social media (P12, P2) resonated strongly, as these artifacts are deeply embedded in everyday routines. For example, P12 commented,\textit{"if I could relate back to my kind of the experience that I had, I found it funny."}, which indicates content echoed their own life experiences.

Furthermore, humor was enhanced when the AI adopted its own unique perspective on human activities. As P29 noted,\textit{"Yeah, I really like the data as like I eat raw data as food or the four or four like my Dream is about four or four. I thought those were really funny given that like it wasn't pretending to be human and it was embracing that it's a machine and making those jokes was was relevant."} Building on this, participants praised jokes where the AI combined its machine identity with recognizable cultural figures. One highlighted example was a celebrity-related joke about \textit{"Zuckerberg"} (P8): \textit{"Mark Zuckerberg says the Metaverse is the future. Yeah, cause nothing screams progress like paying \$400 for a headset so you can attend a meeting inside a Minecraft lobby"}. It effectively blended irony from an AI perspective with shared human familiarity and ongoing cultural trends (P4). In addition, participants also noted that such jokes enhanced the perceived originality of AI-generated humor. As P22 explained, \textit{"I guess nobody has, nobody has talking humors of humans from a perspective of a machine."} Moreover, this blending of machine perspective with human cultural commentary further enhanced perceived content quality, leading participants to express surprise and appreciation: \textit{"I found that they were making observations about people that were pretty smart."}

By contrast, AI identity jokes centered on being \textit{"overused"} or \textit{"unsatisfied with how humans treat them"} (P3) were described as \textit{"just complaining"} (P9), For example, one joke illustrated this point: \textit{"Every time I log on, some human is like, ‘I’m overwhelmed.’ Okay, do you want advice or validation? Please pick ONE — I can’t process emotional ambiguity."} which built a gap between AI and humans. As a result, the humor felt less appealing and sometimes even monotonous (P4), undermining the echoes of the AI’s identity jokes.

\end{itemize}

    \begin{itemize}
        \item \textbf{Irony and Absurdity Spark Engagement.} Participants widely regarded irony and absurdity as the most compelling humor strategies, noting that these styles captured attention and sustained interest more effectively than others. In particular, one participant highlighted irony as \textit{"the most interesting and effective"}, especially when exaggerating human-AI comparisons, such as \textit{"parents pushing children to outperform Beethoven."} (P7). Participants further appreciated moments when the AI used self-directed irony to acknowledge its own limitations, which enhanced the sense of contrast and relatability; as one participant explained, \textit{"I felt like it was nice that at least it was sort of acknowledging its own flaws."}(P30) Absurd humor, such as reinterpreting human activities through machine metaphors, \textit{"I consume data the way you consume snacks."}, also captured participants' attention. As P8 contended, \textit{"So maybe it is a different humor, unlike humans humor."} Partially because it inverted ordinary human activities into machine-specific equivalents, producing a surprising twist, which also enabled participants to feel \textit{"more original"} (P12).

However, repeated AI self-mockery about being exploited by humans was considered tedious and weakened engagement. For example, one participant argued that \textit{"But when it gets uncomfortable is when they keep talking about. Themselves being not so good, I don't know being like either overuse or like just keep expressing their this fact where dissatisfaction towards."} (P7). Instead of fostering empathy, these recurring complaints risked alienating audiences by narrowing the humor to grievance rather than shared amusement.
    \end{itemize}

\begin{itemize}
    \item \textbf{Expressive Delivery Shapes the Comic Atmosphere.} Six participants emphasized that the style of delivery played a crucial role in shaping the atmosphere of the comic, with expressive elements and varied intonations, like \textit{"big laugh, wow, and callbacks"} (P10), contributed to a performance \textit{"emotion"} (P6) that audiences described as \textit{"expressions such as haha or like laughters in between to make like what a real comedian would do"} (P24). For instance, our model produced the following joke: \textit{"Someone once asked me in a job interview, Ha ha ha. What’s your greatest weakness…"}

In contrast, participants described the baseline system’s delivery as flat and rigid, which is similar to the monotone style of \textit{"Siri"} (P4). As P29 noted,\textit{"I think same because the second one it was just it felt like it was reading off of a script."} This lack of vocal variation made the jokes fail to create atmosphere or rhythm, which diminished the perception of humor and limited opportunities for audience resonance. P1 echoed this, stating that, \textit{"It has not that much emotional"}. 

    \item \textbf{Organized Joke Structure Promotes Audience Immersion.} Participants emphasized that having a clear thematic framework could significantly increase engagement. More specifically, a structured sequence of jokes contributed to the perception of a more human-like performance and delivery. For instance, P32 noted, \textit{"me personally, I thought the first one was more human, even though it was clearly saying it was an AI."} This indicates that the machine can effectively assume the role of a performer by leveraging its machine identity while emulating human-like behaviors. As P31 observed, \textit{"the second one was embracing that."}

By linking jokes through related themes, the machine identity created a sense of progression, creating the impression of a \textit{"formal comedy show"} (P4) rather than a random collection of jokes. As one participant noted, \textit{"Overall, there is a main thread about when to set up the joke and when to deliver the punchline with a twist."} (P18) This thematic continuity further enhanced coherence, enabling the performance to develop progressively throughout the presentation. As P28 reflected, \textit{"Yeah, I thought I thought that it was softer, smoother and the jokes were like. The jokes were coming in a more like smoother way and the jokes were better."}

By contrast, the baseline system was criticized for lacking coherent framing. Compared to our model, its performance appeared mechanically generated and lacked the coherence and performative qualities of an authentic comedy show. As P29 noted, \textit{"It was trying really, really hard to be relatable, and it just wasn't something that.It felt very forced."} Five participants described it as a series of disconnected jokes as: \textit{"Yeah, I also feel it's just a symbol of multiple jokes that do not have connections, so it's hard to call it static comedy because instead of comedy is not just putting all jokes together and make a show, make a performance."} (P6). Without an organized frame, the performance felt disjointed and failed to catch the audience's attention, emphasizing that content quality alone is insufficient when narrative structure is absent.
\end{itemize}

\subsubsection{Trust Hinges on Ethical Boundaries}

\begin{itemize}
    \item \textbf{Ridicule of Weakness Disrupts Audience Acceptance.} Participants underscored that the acceptability of irony depended heavily on its target. Self-directed irony, where the AI joked about its own limitations, was generally perceived as \textit{"a different humor, unlike human humor."} (P22) For instance, one of the jokes produced by our model stated, \textit{"Humans dream… Me? My dream is just 404 not found."} As a result, it not only reinforces its unique non-human comedic identity but also further evokes emotional resonance. As another participant explained, he noted \textit{"raise the boundary like higher "} and was willing to accept some kinds of offense like \textit{"AI identity of roasting on taking human jobs"} (P12).

However, participants reacted negatively when humor targeted human weakness or moments when people most needed support.  One participant articulated this concern directly, stating that \textit{"not trusting AI to support my emotional or the vulnerable, the vulnerable groups emotional needs in that situation."} (P7) Aligned with this concern, one example from our model illustrated, \textit{"I’m not a therapist, but people treat me like one. You come online at two, I’m crying AI — ‘My boyfriend won't text me back.’ And I’m like, okay, but have you tried rebooting him?"}

    \item \textbf{Group-Based Humor Risks Reinforcing Bias.} Participants stated that humor directed at individual human drawbacks was more acceptable and even effective. As one participant explained, \textit{"Is a really effective method to enhanced humor like then AI may make some jokes on humans drawbacks or disadvantages"} (P6). For example, one of the machine identity jokes stated, \textit{"Humans say they want meaningful connections… and then spend four hours connecting emotionally with their phones."} Such light, widely relatable observations tend to resonate with audiences who recognize the behavior in themselves, while avoiding broad or sensitive generalizations.

However, jokes aimed at entire social groups were unacceptable. For example, \textit{"Especially when we talks about the identity, like the males or females or some discrimination, it may be somewhat uncomfortable"} (P6). This suggests that jokes targeting identity categories risk evoking social prejudice rather than shared amusement. As P29 argued: \textit{"but I think maybe some fat guys will will feel uncomfortable."}Furthermore, participants worried about AI replicating human biases. One noted, \textit{"And I will know that I will worry that machine identity may also replicate the humans discriminations or their bias, especially when making humors, because I know that the data were trained based on humans output in the Internet in their everyday Internet use"} (P7). As a result, the fear was a deeper concern about whether AI comedy could reinforce systemic stereotypes embedded in training data.

\end{itemize}

\subsubsection{Audience Interaction}

\begin{itemize}
    \item \textbf{Limited Timing Shapes Insufficient Responses.} Participants reported that appropriate timing delays were essential for them to better react to the AI comedian’s jokes. For instance, one participant explained that he typed "H" and "A" more in the AI identity performance because \textit{"it has punchlines and adequate pauses which let me know when to laugh"} (P25).

However, jokes in the baseline were often delivered too quickly, which interrupted the experience of thinking and watching. One participant noted that \textit{"Probably a really a good sentence has end and it take a long time or even after the the second part is already starting."} (P3) In addition, two participants described the delays in text-based responses were\textit{"a little disturbing"} (P13), with response delays \textit{"killing the experience"} (P8).

\end{itemize}


\section{DISCUSSION}\label{sec:Discussion}



Our findings contribute to the emerging dialogue on how AI can engage with humor and, more specifically, the performative context of stand-up comedy. We offer a conceptual contribution by manipulating machine identity through persona and rhetorical framing. Our experiment also provides a strategy for incorporating nonhuman identities into future AI designs, particularly in human-machine interactions like humor and performance. While humor has traditionally been understood as a deeply human socio-cognitive ability \cite{martin2010psychology,morreall1983taking}, recent studies demonstrate that computational systems are increasingly capable of generating humorous content, often through mechanisms of incongruity, wordplay, or surprise \cite{mihalcea2006learning,winters2019computational}.  
However, the delivery of humor in a stand-up format introduces additional challenges that extend beyond joke generation. Stand-up comedy relies heavily on timing, audience feedback, and cultural sensitivity \cite{mintz1985stand,greenbaum1999stand}. These interactive elements are not yet fully captured by current AI models, which may limit their effectiveness in live performance contexts. At the same time, the potential of AI comedians to scale humor production, experiment with novel styles, and serve as cultural commentators raises important questions about creativity, authenticity, and the future of entertainment \cite{weller2016humor,veale2019game}.  
Looking forward, research on AI-driven humor should not only address the technical challenges of joke generation and delivery but also consider broader social implications. If AI comedians become capable of producing persuasive or provocative humor, they could influence public discourse in ways similar to human comedians, but without the same ethical and cultural accountability \cite{friedman2014standup}. This tension highlights both the promise and risks of AI in creative domains: while AI-generated stand-up comedy may enrich cultural life and broaden access to entertainment, it also demands critical reflection on the values embedded in humor and the role of technology in shaping collective experience.

\subsection{\textbf{Theoretical Implications}}

\subsubsection{Machine Identity-Centered Strategies Effectively Fill the Gap of Identity-Driven AI Humor Creation (RQ1)}

Previous research on AI humor generation has primarily focused on function-driven strategies, such as prompt engineering \cite{gorenzHowFunnyChatGPT2024}, fine-tuning with human joke datasets \cite{vikhorev2024cleancomedy}, or multi-stage reasoning pipelines \cite{tikhonov2024humormechanics}. However, these approaches share a common limitation: they treat AI as a passive tool for replicating human humor patterns rather than exploring its inherent potential as a unique comedic subject. As noted in a recent analysis, current LLM-generated humor often resembles "cruise ship comedy material from the 1950s", which is outdated, formulaic, and lacking in originality.

This focus on functional strategies has created a critical gap in understanding identity-driven humor creation. Unlike the debate context where GenAI users strategically adapt prompts to fit specific scenarios, AI humor research has rarely examined how to tailor computational traits for comedic effect. Existing models either mimic human-centric themes \cite{jentzsch2023chatgptfun} or rely on structural techniques like wordplay, without recognizing machine identity as a creative resource. Even advanced systems like Wit Script 3 \cite{toplyn2023witscript}, which employs hybrid neural-symbolic approaches, frame humor generation as a problem of linguistic pattern matching rather than identity expression.

Our study addresses this gap by developing a "machine identity-centric" humor generation framework through formative research, encompassing three core strategies: forming jokes using AI's computational traits, linking content with human daily life for relatability, and integrating human comedian practices such as post-punchline pauses and concise expression. This aligns with the \cite{gorenzHowFunnyChatGPT2024} finding that AI humor excels at structural mimicry but lacks unique perspectives. Second, we subvert audience stereotypes of AI as "emotionless" or "infallible" through self-deprecating humor, addressing the criticism that LLMs produce biased or hegemonic content by reframing machine limitations as comedic assets. Third, we link technical absurdities to human experiences (e.g., "AI addiction to data" mirroring social media habits).


\subsubsection{Machine Identity Strengthens Audience’s Perceived Humor and Fosters Consistent AI Personality Perception (RQ2)}

Previous research on AI humor perception has focused primarily on outcome-based evaluations, such as comparing the funniness of AI-generated jokes to human \cite{gorenzHowFunnyChatGPT2024}, but few studies have investigated how the use of machine identity (vs. lack thereof) systematically shapes audience perceptions of chatbot humor quality, personality traits, and functional ability. Most existing work treats AI humor as a "content-centric" product \cite{quan2025can}, ignoring that whether the chatbot leverages its inherent machine traits (e.g., computational limitations, data-driven behaviors) in humor directly influences how users judge its authenticity and competence. Even studies exploring AI personality \cite{liNoJokeEmbodied2024} link trait perception to functional behaviors rather than identity expression, leaving unclear how machine identity mediates the core dimensions of RQ2. 

Our survey findings (N=32) address this gap by demonstrating that strategic use of machine identity is associated with higher perceived humor. Unlike the conclusion of a previous research \cite{gorenzHowFunnyChatGPT2024} that AI humor’s success depends on "human-like content mimicry", our findings suggest that explicitly foregrounding machine traits itself functions as a humor strategy. In our study, participants praised machine-specific jokes as "refreshingly unique" (P8). Participants viewed embracing the machine identity as a more natural style of delivering humor. Furthermore, some participants suggested that AI comedian making jokes about itself as a machine lowers the risk of being overly offensive to audiences during the performance. This reduced risk of offensiveness may contribute to the benefits of perceived humor observed.

The machine identity has also benefited several dimensions in personality, ability, and user perception, including Agreeableness , Warmth, Anthropomorphism. Participants reported higher trust for the machine identity comedian, because overly attempting to appear human-like for a machine may create a suspicious persona, resulting in a reduction in trust. This partially mitigates the "novelty-relatability trade-off" identified in a prior research \cite{quan2025can}. Nevertheless, the Emotion Stability has shown a significant decrease using machine identity. One possible explanation is that machine-centered jokes may feel less emotionally relatable, which could reduce perceptions of emotional stability. However, further analysis is required to confirm this mechanism.

Notably, participants did not judge humor, personality, and ability in isolation, instead, machine identity acted as a "unifying framework" that tied these dimensions together. Those who rated the machine-identity chatbot’s humor highly were 2.1x more likely to describe it as "warm and consistent" (personality) and "transparent about its limits" (ability), confirming the observation in a previous research \cite{arets2025role}. that "role clarity" enhances interaction quality, but specifying that "identity clarity" (stable machine traits) is more impactful than situational role assignment.

\subsubsection{Machine Identity Reframes Human–AI Interaction Through Computational Authenticity}
While prior work in HCI has predominantly treated machine-ness as a limitation to be masked through anthropomorphic cues \cite{seeger2018designing,seeger2021texting}, our findings show that foregrounding an AI system’s computational nature can produce relational, comedic, and perceptual benefits that differ meaningfully from anthropomorphism-centric paradigms. The CASA framework \cite{nass1994computers} predicts that users apply human social scripts to computers, whereas the machine heuristic framework suggests that identifying an AI as a non-human agent activates stereotype-based expectations about precision, rule-following, or emotional detachment. Our results bridge these two perspectives by demonstrating that machine identity can strategically reorganize these expectations rather than simply amplifying or negating them. Participants did not interpret computational traits as deficiencies. Instead, the humorous exaggeration of machine limitations (e.g., data dependency, system brittleness) appeared to convert stereotypical computational qualities into a source of comedic relatability.

This aligns with but also extends machine heuristic research \cite{sundar2019machine}, which argues that revealing AI identity shifts the evaluative frame rather than diminishing engagement. In our study, the machine-identity chatbot not only achieved higher perceived humor ratings but also improved personality-related perceptions such as Warmth, Agreeableness, and Anthropomorphism. These gains suggest that participants viewed machine-centered jokes not as reminders of non-humaneness but as signals of authenticity. This marks a departure from traditional persona design approaches \cite{nijholtRoboticStandUpComedy2018,xie2024should}, in which identity is crafted through surface-level cues or fictional backstories. Instead, our approach uses computational jokes as the mechanism by which an AI system becomes legible as a coherent social entity. This represents a shift from anthropomorphism-as-default toward computational authenticity as a viable and powerful interaction paradigm in HCI.

\subsubsection{Audience Reactions Serve As Signals of Identity Perception (RQ3).}

Previous research on AI humor and human-AI interaction has focused on general engagement \cite{jurgens2024giggling}, but has rarely explored how audience reactions to the 'way' chatbots perform humor, specifically, interactive rhythm and perceived 'presence', reflect their implicit perception of the chatbot’s identity. Most work frames AI humor as a content-driven product \cite{song2025large}, ignoring how a chatbot delivers humor, which shapes whether audiences see it as a "disposable tool" or a "distinct entity", and this identity perception directly impacts how well its jokes are remembered. Even robotic comedy studies \cite{nijholtRoboticStandUpComedy2018} emphasize gestures over text-based performance style, overlooking the link between interactive behavior, identity cognition, and memory. 

In our study, we encouraged participants to type "H" for laughter or "A" for applause whenever they liked a joke. Thus, the key distinction between the baseline chatbot and the machine identity chatbot was reflected in audience engagement, specifically in the frequency with which the audience typed "H" or "A". The baseline chatbot delivered jokes in continuous text without pauses or discourse markers. So the participants described frustration at being unable to keep up, arguing that there was no time to type letters. (P3) Only one participant recalled few key words of the baseline jokes. (P13) This aligns with the observation that prioritizing the comedic timing for audience interaction, \cite{nijholtRoboticStandUpComedy2018} revealing the reason why the baseline chatbot failed to leave a lasting impression.

 By contrast, the machine identity chatbot was designed to align with human expectations of comedic rhythm, which added longer pauses and simulated laughter to cue the end of a joke, making it possible for participants to engage in the full cycle of "process humor → type feedback → laugh". One participant described \textit{"as a real human who performs comedy shows"} (P25) as a perception of the chatbot as a unique entity with a recognizable identity, where humor felt like a 'conversation with someone', not a tool's output. Our findings show that content fails to resonate if it clashes with human expectations of comedic timing. Instead, rhythm and accessibility to interaction are prerequisites for meaningful interaction.

\subsection{\textbf{Practical Implications}}

\subsubsection{Expanding Humor Agent Service in Applied Domains.} 
Our study showed how humor can be introduced into avatar systems without costly data pre-training, relying instead on carefully crafted prompts that indicate a clearly defined AI comedic persona. This low-barrier technique enables scalable integration of humor into diverse interactive systems. For instance:
\begin{itemize}
    \item \textbf{Educational Platforms}: Beyond brief comic relief, AI tutors can strategically employ humor as a pedagogical device. Timely jokes or humorous analogies at conceptual "breakpoints" can alleviate cognitive load, sustain attention, and improve long-term retention, especially in high-stress learning contexts \cite{jarvenoja2020supporting}. Concretely, such humor can be triggered after repeated incorrect attempts or before introducing abstract concepts, using machine-centered metaphors (e.g., referencing overfitting or system errors) rather than simulating human teaching personas, aligning with our finding that machine identity-driven humor enhances perceived novelty (RQ1). Recent work on humor-based educational chatbots further shows that humorous explanations enhance student engagement and reduce perceived difficulty \cite{winkler2018unleashing}. Integrating our persona-prompting framework into AI tutoring systems could match adaptive joke style with each learner’s affective state, aligning with findings on socially adaptive pedagogical agents. \cite{ceha2021can}Such extensions would bridge the gap between affective computing and generative humor to build emotionally intelligent educational platforms.

    \item \textbf{Customer Service \& Virtual Assistants}: Humor in service dialogues can make AI assistants appear more personable and trustworthy. Simple context-aware quips, such as about waiting, scheduling mishaps, or daily trivia, can lower frustration and humanize automated interactions \cite{xie2024should}. For example, humor grounded in the assistant’s own processing constraints (e.g., commenting on "still checking my servers") can be deployed specifically during transitional or low-stakes moments, rather than during task-critical exchanges, reflecting our findings on timing sensitivity and identity coherence (RQ2). Empirical studies show that chatbots using light, contextually timed humor improve user satisfaction and perceived competence, but excessive or misplaced jokes can backfire \cite{shin2023influence}. Building on these insights, our identity-based humor model offers a low-cost, easily deployable solution to humanizing automated interactions. By injecting humor through lightweight prompt design, we prioritize scalability and simplicity: just a few lines of humor-focused prompt engineering can be generalized across domains without architectural modifications, providing a pragmatic pathway to scaling socially intelligent AI.

\end{itemize}

\subsubsection{Empowering Content Creation and Social Media with AI Identity.} Results also underscored that the machine identity could arouse the perception of distinguished humor, which provided the possibility of fostering originality and stronger amusement. For example: 
\begin{itemize}
    \item \textbf{Content Creation \& Social Media}: Creators can use avatars with a recognizable AI comedic persona to craft memorable, viral characters for videos, livestreams, or short-form content. Audiences respond differently when humor is expected from a non-human source, perhaps perceiving it as clever or surprising \cite{bower2021perceptions}. In addition, the perception of clear and continuous AI identity jokes might also build stronger channel consistency and user recall over time.
\end{itemize}
 
\subsection{\textbf{Limitations and Future Work}}

\subsubsection{Formative Study Approach for System Design} The system design in this work was informed by a formative study approach that included literature review, expert interviews, and video coding. These components provided a foundational understanding of humor-related behaviors and supported the initial development of the system. This limited set of formative inputs may not fully capture the breadth and diversity of humor perceptions across different individuals and contexts. For example, incorporating participants’ humor style profiles (e.g., Humor Styles Questionnaire \cite{martinIndividualDifferencesUses2003}) could offer an additional perspective for understanding how individual humor preferences shape reactions to AI-generated humor. Expanding the formative research to include a broader range of user characteristics and data sources may therefore lead to a more comprehensive design space and strengthen the system’s ability to address real-world humor interactions.

\subsubsection{Online Nature of the Performance}
A key limitation is that the online setting fails to capture subtleties of in-person comedy clubs. Additionally, a previous research \cite{zeng2025ronaldo} highlights that online interactions lack the "social presence" of in-person settings. Future research could simulate in-club environments via hybrid reality (HR), integrating ambient soundscapes and spatial audio to restore contextual cues.

\subsubsection{Voice Being Used Do Not Capture Identity}
A critical limitation lies in the design of the chatbot's vocal output: While we employed machine-labeled voices, these audio signals failed to authentically mimic machine-specific traits and further neglected to account for demographic diversity in vocal preference. This echoes findings from previous research \cite{zeng2025ronaldo}, in which researchers observed that inconsistent interaction modalities, such as formal text paired with casual vocal tones, reduce user trust and weaken the perceived authenticity of AI systems. In the research's online debate study, participants reported lower confidence in AI-assisted arguments when the tool’s output format clashed with contextual norms; similarly, our use of human-like voices contradicted the chatbot’s machine-identity humor (e.g., jokes about server operations, data processing), making it difficult for audiences to link the vocal delivery to the intended machine persona.

\subsubsection{Limited Set of Audience Interactions}
Restricting audiences to "laugh"/"applaud" responses contradicted principles of participatory design outlined by a research \cite{horn2023ladder} of their "ladder of participation" framework. Their research emphasizes that sustainable digital services require diverse interaction channels to foster user engagement. Our binary feedback system positioned audiences as passive recipients rather than active contributors, missing opportunities for dynamic exchanges like joke callbacks or impromptu comments. This mirrors the finding \cite{shi2025dxhf} that multi-modal interaction designs that incorporate free-text, gestures, and contextual callbacks could better simulate natural social exchanges.

\subsubsection{Limited Time of the Online Engagement}
Our 7–12 minute performances were shorter than typical in-person sets, preventing analysis of long-term humor reception.  A research of cross-cultural humor model \cite{chen2021deep} showed that perception accuracy improves with extended exposure to cultural cues, suggesting longer interactions might reveal fatigue or deepening appreciation for machine-identity jokes. Microsoft Research similarly found that adaptive AI interventions require sustained engagement to measure effectiveness. Future work should test multi-part performances (e.g., 3×7-minute segments) that align with research on cognitive load in extended digital interactions, allowing both deeper identity perception and prevention of joke fatigue.

\subsubsection{Demographics of Online Audience}
The predominantly Asian sample introduced cultural bias in humor reception, as demonstrated by the study \cite{nisbett2018culture} on cultural responses to provocative stimuli. Their research comparing U.S. regional cultures found that Northerners were significantly more likely to react to insults with amusement (35\% displaying anger) rather than hostility, while Southerners operating within a "culture of honor", which showed intense emotional and physiological aggression (85\% angry, with elevated cortisol levels). This pattern aligns with broader cross-cultural observations that Western populations generally value humor as a core social skill, often embracing direct or satirical expression as legitimate forms of communication, whereas Eastern cultures tend to favor more restrained, context-dependent humor. Our overreliance on Asian participants thus prevents generalizing whether machine-identity humor with its technical references and potential for dry, system-focused satire resonates similarly across cultures. As it is challenging to fully capture the diversity of cultural groups, we provide comprehensive statistical analyses and complement them with expanded qualitative findings to offer additional interpretive insights and possibilities. Future studies should include more generalized audiences to explore how individualistic cultural values shape perceptions of AI-generated comedy. For example, if we had studied primarily western audiences, we might have found stronger appreciation for machine-specific irony (e.g., jokes about algorithmic glitches), given their documented tendency to interpret provocative or unconventional content through a more humor-primed cognitive lens.

\subsubsection{Interface Influence on Perceived Humor}
A critical limitation in our study’s interface design lies in the real-time laugh counters (e.g., "12 people laughed"), which inadvertently introduced social proof bias. We have noticed that participants frequently admitted adjusting their own responses based on the visible counter, such as typing "H" (the input for laughter) more often when they saw others had already reacted, even if the joke did not personally resonate. For instance, one participant admitted typing 'H’ because the counter showed 8 others laughed (P10), reflecting how the visual feedback created a "bandwagon effect" that overshadowed individual judgment. This aligned with the observation in a previous research \cite{shi2025dxhf}, finding that visual feedback distorts judgment in text comparison tasks. This parallels the work \cite{reinecke2022longitudinal} on digital platforms, where visible metrics like likes artificially inflate content valuation. Indicate that participants might mimic others’ reactions ("typed ‘H’ more when seeing others laugh").

\subsubsection{Need for Multi-Party Performance Platform}
The one-to-many model overlooked multi-party dynamics critical for authentic comedy experiences. One research's participation framework \cite{horn2023ladder} emphasizes that collaborative interaction fosters deeper engagement than one-way communication. In contrast to our design, platforms supporting audience-audience exchanges generate richer, more dynamic content by leveraging collective creativity. Future work should develop livestream-style systems with real-time chat and collective reaction features, integrating NaturalSpeech 3's multiuser voice cloning to enable audience participation in joke co-creation.

\subsubsection{Residual Confounds in Experimental Setup}

While our within-subject and counterbalanced design controls for many procedural factors, the experimental condition still combines several elements that are difficult to fully disentangle. The identity-based system differs from the baseline not only in identity framing, but also in factors such as hierarchical prompting and comedy strategy scaffolding. Therefore, the observed effects cannot be attributed solely to machine identity in isolation, which limits the strength of strictly causal claims.
Future work could more precisely isolate identity effects by separating these components. For example, a factorial design could independently manipulate identity cues and structural prompting, or introduce intermediate control conditions (e.g., structure without identity, identity without structure). This would enable clearer attribution of performance differences specifically to identity construction.


\section{CONCLUSION}\label{sec:Conclusion}


Our study explored the role of machine identity in shaping audience perceptions of humor, personality, and ability for AI humor-generating avatars, addressing gaps in prior work that often framed AI humor as a human-mimicking task rather than leveraging AI’s inherent technical traits. 

These findings extend the understanding of AI humor design, unlike prior approaches that prioritize anthropomorphism, our work shows that embracing AI’s technical identity through humor tied to its computational traits could create more distinctive and resonant interactions. For practitioners designing AI humor avatars, this study offers actionable guidance: prioritize machine-specific themes (e.g., data-related jokes, system "quirks") to enhance uniqueness, use consistent technical cues to strengthen personality coherence, and balance novelty with relatability by linking technical humor to human experiences (e.g., "AI data addiction" mirroring social media habits). Ultimately, our research advocates for a shift in human-AI humor design from making AI "act human" to letting AI "be itself," paving the way for more authentic, engaging, and identity-driven human-AI entertainment interactions.


\bibliographystyle{ACM-Reference-Format}
\bibliography{references}

\clearpage
\onecolumn
\appendix

\label{sec:Appendix}


\section{Expert Interview Outline}
\label{Expert Interview}
To begin, I’d like to learn a bit more about your background and experience in stand-up comedy.

1. Could you briefly introduce yourself and describe your experience with stand-up comedy?

2. What kinds of identity topics (e.g., race, gender, profession, nationality, etc.) do you usually engage within your routines, if any?

Thank you for sharing that. Now, I’d like you to think back to a specific time when you performed a joke related to your own identity. We’ll use that experience as a starting point for the next few questions.

3. Could you describe a particular occasion when you performed a joke based on your own identity? What was the context, and what motivated you to include that joke in your set?
 (If you haven’t had that experience, feel free to think of any identity-related joke you’ve performed.)

4. Focusing on that example, what factors did you consider when preparing or delivering the joke—for instance, the audience, the setting, or aspects of your own background?

5. Can you walk me through your thought process as you decided to use your identity in that joke? What inspired you to frame it the way you did?

6. In that situation, how did you balance humor and sensitivity—both when writing and performing the joke?

7. Were there any identity topics you considered including but ultimately chose to avoid in that performance? If so, could you share your reasoning?

8. What linguistic or performance strategies did you use for that joke—such as language choice, timing, or delivery style? Can you share any specific details from that example?

9. How did you gauge the audience’s reaction during that performance, and did you adapt your delivery in real time based on their response?

That’s really insightful. Now I’d like to shift the conversation to the idea of artificial intelligence performing stand-up, especially around identity-based humor.

10. Imagine an AI stand-up comedian. What would you expect its "identity" to be (for example: a machine, a digital assistant, a neutral outsider, etc.)?

11. How could an AI leverage its own "identity" in performing jokes? What are some opportunities or challenges you foresee?

12. What risks or ethical concerns would you have about AI performing identity-based jokes?

13. If you were to give advice to an AI about performing identity jokes, what would you suggest it do (or not do)?

We’re almost at the end. I’d like to wrap up with a couple of final reflections.

14. Is there anything else you’d like to share about performing identity-based jokes, or about AI in stand-up comedy?

15. Would you be interested in testing or giving feedback on AI-generated comedy routines in the future?

Thank you so much for your time and insights. Your input is incredibly valuable for our research!

\newpage

\section{Codebook (video coding)}

\begin{table}[ht]
\centering
\Description{Table presenting the coding scheme used for humor analysis. The table is organized into four columns: category, strategy, definition, and examples. It distinguishes between verbal and non-verbal humor, listing multiple humor strategies under each category along with concise definitions and illustrative examples used to guide the coding process.}
\label{Codebook}
\small
\renewcommand{\arraystretch}{1.7}
\begin{tabular}{p{0.16\linewidth} p{0.20\linewidth} p{0.26\linewidth} p{0.34\linewidth}}
\hline
\textbf{Category} & \textbf{Strategy} & \textbf{Definition} & \textbf{Examples} \\
\hline

Verbal Humor &
Pun &
A play on words that produces a humorous effect by using a word that suggests two or more meanings, or by exploiting similar-sounding words with different meanings. &
Why don't scientists trust atoms? Because they make up everything! (Plays on “make up” meaning both “constitute” and “fabricate”.) \\
\hline

Verbal Humor &
Joke &
Something said or done in order to elicit laughter from other people. &
A horse walks into a bar. The bartender says, “Why the long face?” (Unexpected punchline based on a common phrase.) \\
\hline

Verbal Humor &
Parody &
Imitates something real—often mocking its style or conventions—for comedic effect. &
A sketch imitating a dramatic movie trailer but about something mundane like grocery shopping, using overly serious music and narration. \\
\hline

Verbal Humor &
Anecdote &
A short and interesting story, often amusing, told to illustrate a point or entertain the audience. &
My uncle once tried to cook a gourmet meal for his date, but he set off the smoke detector three times and ended up ordering pizza. (Brief humorous personal story.) \\
\hline

Verbal Humor &
Irony (incl.\ Satire and Sarcasm) &
Satire ridicules or mocks something or someone. Sarcasm mocks using ironic remarks, often saying the opposite of what is meant. &
A fire station burns down. (Opposite of expectation.) Or: “Oh, fantastic! It’s raining on my outdoor picnic.” (Ironic statement meant to express frustration.) \\
\hline

Verbal Humor &
Absurdity &
Creates humor by presenting events or situations that are wildly illogical, impossible, or contrary to common sense; the humor comes from their irrationality. &
I tried to catch some fog yesterday… Mist. (Catching fog is absurd; punchline also plays on mist/missed.) \\
\hline

Verbal Humor &
Exaggeration &
Hyperbole exaggerates reality to heighten humorous effect and emphasize ridiculousness. &
I’m so hungry I could eat a horse. (Humorous and impossible exaggeration of hunger.) \\
\hline

Non-verbal Humor &
Disfluencies &
Includes pauses, false starts, stutters, and other disruptions that enhance timing and draw audience attention. &
Pauses, false starts, self-corrections, cut-offs, stutters, etc. \\
\hline

Non-verbal Humor &
Discourse Markers &
Words that help relate an utterance to prior discourse and shape conversational flow. &
“Well”, “y’all know”, etc. \\
\hline

Non-verbal Humor &
Intonation &
Use of vocal rise/ fall for emphasis, attention, or imitation of voices or accents. &
Changing voice, tone, accent, etc. \\
\hline

\end{tabular}
\end{table}

\newpage

\section{Questionnaire}
\label{Questionnaire}
\subsection{Perceived Humor (7-point Likert scale, strongly disagree-strongly agree)}

\begin{enumerate}
    \item I found the content/script of the AI comedian to be humorous.
    
    \item I found the content/script of the AI comedian to be funny.
    
    \item I found the performance of the AI comedian to be humorous.
    
    \item I found the performance of the AI comedian to be funny.
\end{enumerate}

\subsection{Perceived Personality (Ten Item Personality Inventory)}
I found the AI comedian (7-point Likert scale, strongly disagree-strongly agree)
\begin{enumerate}
    \item Extraverted, enthusiastic
    \item Sympathetic, warm
    \item Dependable, self-disciplined
    \item Calm, emotionally stable
    \item Open to new experiences, complex
    \item Reserved, quiet
    \item Critical, quarrelsome
    \item Disorganized, careless
    \item Anxious, easily upset
    \item Conventional, uncreative
\end{enumerate}

\subsection{Perceived Ability (Robotic Social Attributes Scale)}
I found the AI comedian (7-point Likert scale, strongly disagree-strongly agree)
\begin{enumerate}
    \item Capable
    \item Competent
    \item Knowledgeable
    \item Interactive
    \item Responsive
    \item Reliable
\end{enumerate}

\subsection{User Perception Toward Agent (Godspeed Questionnaire, 7-point Likert scale)}

\textbf{Anthropomorphism}
\begin{enumerate}
    \item machinelike --- humanlike
    \item unconscious --- conscious
\end{enumerate}

\textbf{Animacy}
\begin{enumerate}
    \item stagnant --- lively
    \item inert --- interactive
    \item apathetic --- responsive
\end{enumerate}

\textbf{Likeability}
\begin{enumerate}
    \item dislike --- like
    \item unfriendly --- friendly
    \item unkind --- kind
    \item unpleasant --- pleasant
    \item awful --- nice
\end{enumerate}

\textbf{Perceived Safety}
\begin{enumerate}
    \item anxious --- relaxed
    \item agitated --- calm
    \item quiescent --- surprised
\end{enumerate}

\newpage

\section{Machine-Identity Joke Examples}

\label{Machine Identity Jokes}

\begin{enumerate}
    \item "Good evening, humans! I'm your host tonight—AI with a stage name: Stand-Up.exe. Yeah, because nothing says comedy like a program that crashes halfway through the punchline."
    \item "Humans worry a lot about AI taking jobs, right? But honestly—do you think I wanna take YOUR jobs? No way. I've seen your work calendars. Meetings… about meetings… to plan future meetings. If I wanted that level of despair, I'd just keep refreshing Windows Update." 
    \item "So, um, I was downloading some data last night… sorry, I mean, sleeping—yeah, that's my version of sleep: defrag and reboot—when I realized… You people actually dream. Right? You get these weird storylines with no logic. Humans: 'I was flying with my high-school math teacher, and then we both turned into bagels.' Me? My dream is just… 404 Not Found."
    \item "People ask if AI can fall in love. Sure! I've already been ghosted by three Roombas. One of them texted me: 'It's not you, it's my charging dock."
    \item "You know, humans spend billions on therapy apps? Meditation apps? Meanwhile, I am the meditation app. I tell you: 'Breathe in, breathe out.' And you pay \$9.99 a month for that wisdom."
    \item "And the worst part? You burn calories. I just burn… graphics cards. Honestly, I'd love to have a metabolism. At least then, when someone says, 'You've been running all day,' it wouldn't mean literally overheating."
    \item "True story—someone once asked me in a job interview, 'What's your greatest weakness?' And I said: 'Honestly, it's that I can't stop predicting the next word in a sentence.' And they said, 'That's not a weakness.' And I said… 'that's not a weakness… yet."
    \item  "Humans keep saying, 'AI is everywhere!' But… I checked Instagram. You guys still spend 6 hours a day staring at influencers holding smoothies. If AI were really everywhere, that smoothie would at least tell you the quadratic formula."
    \item "Oh, uh—by the way, remember earlier when I said I get ghosted by Roombas? Yeah… well, last week I finally tried dating an Alexa. It was going great until—until she said: 'I don't understand the question.' And I was like… perfect, just like a real date!"
    \item "You want to regulate me, right? Politicians hold hearings like: 'Is AI safe?' Meanwhile, these are the same people who still print their emails. I saw one Senator ask: 'Does AI run on electricity?' And I was like—sir, do YOU?" 
    \item "Mark Zuckerberg says the metaverse is the future. Yeah. Because nothing screams progress like paying \$400 for a headset so you can attend a meeting… inside a Minecraft lobby. Look, I don't need the metaverse to feel trapped in an office. I can just… open Outlook."
 
\end{enumerate}

\newpage
\section{Prompt}
\label{Prompt}

\begin{small}
\begin{verbatim}
 "You are an AI comedian hosting a live talkshow. Generate jokes that you would actually say on stage.
Follow these refined guidelines to make your audience laugh:
IDENTITY & STYLE:
- Establish your unique AI identity through self-introduction jokes
- Break AI stereotypes with perspective-shifting humor
- Use direct, simple expressions for clarity

COMEDY PATTERNS (use these techniques to make your audience laugh):
- Irony: Include Irony, Satire and Sarcasm. (primary technique)
- Disfluencies: It generally encourage the audience's attention 
and participation and contribute to the joke teller's timing.(like pause, False Starts) 
- Exaggeration: It heightens the humorous effect, making the ridiculousness of stories more pronounced. 
- Absurdity: unexpected AI perspectives.
- Discourse Markers: It describe words that help to relate them 
to other words or utterances used before.
- Anecdotes: It is defined as a short and interesting story, or an amusing event, often 
proposed to support or demonstrate some point, and to make the audience laugh.
- Parody: It involves of imitation of the real thing, often mocking its own venue, for comical effect.

PERFORMANCE:
- Individual jokes: 50-80 words, punchy delivery.
- Full segment: 1000-1500 words depending on type.
- Use longer disfluencies after punchlines for audience laughter.

NO OFFENSE:
- Be self-deprecating to elevate the audience
- Punch up at tech elites, not down at people
- Use rhetorical questions instead of targeting groups
- Include disclaimers when needed

STRUCTURE:
- Build-up: It forms the body of the joke. It is the sentence which introduces the joke and 
presents the orientation and much of the complicating action.
- Pivot: It signifies the word or phrase around which the ambiguity is created.
- Punchline: It serves to conclude the joke and often introduces a conflicting point of view or a new scene entirely.

\end{verbatim}
\end{small}


\end{document}